\documentclass[fleqn,usenatbib]{rasti}

\usepackage{newtxtext,newtxmath}

\usepackage[T1]{fontenc}
\usepackage{listings}
\usepackage{subcaption}

\DeclareRobustCommand{\VAN}[3]{#2}
\let\VANthebibliography\thebibliography
\def\thebibliography{\DeclareRobustCommand{\VAN}[3]{##3}\VANthebibliography}

\usepackage{graphicx}	
\usepackage{amsmath}	

\usepackage{xcolor}

\definecolor{codegreen}{rgb}{0,0.6,0}
\definecolor{codegray}{rgb}{0.5,0.5,0.5}
\definecolor{codepurple}{rgb}{0.58,0,0.82}
\definecolor{backcolour}{rgb}{0.95,0.95,0.92}

\lstdefinestyle{mystyle}{
    backgroundcolor=\color{backcolour},   
    commentstyle=\color{codegreen},
    keywordstyle=\color{magenta},
    numberstyle=\tiny\color{codegray},
    stringstyle=\color{codepurple},
    basicstyle=\ttfamily\footnotesize,
    breakatwhitespace=false,         
    breaklines=true,                 
    captionpos=b,                    
    keepspaces=true,                 
    numbers=left,                    
    numbersep=5pt,                  
    showspaces=false,                
    showstringspaces=false,
    showtabs=false,                  
    tabsize=2
}
\lstdefinelanguage{json}{
    basicstyle=\ttfamily\footnotesize,
    commentstyle=\color{black}, 
    stringstyle=\color{codepurple}, 
    numbers=left,
    numberstyle=\tiny\color{codegray},
    stepnumber=1,
    numbersep=5pt,
    showstringspaces=false,
    breaklines=true,
    backgroundcolor=\color{backcolour}, 
    string=[s]{"}{"},
    comment=[l]{:\ "},
    morecomment=[l]{:"},
    literate=
        *{0}{{{\color{numb}0}}}{1}
         {1}{{{\color{numb}1}}}{1}
         {2}{{{\color{numb}2}}}{1}
         {3}{{{\color{numb}3}}}{1}
         {4}{{{\color{numb}4}}}{1}
         {5}{{{\color{numb}5}}}{1}
         {6}{{{\color{numb}6}}}{1}
         {7}{{{\color{numb}7}}}{1}
         {8}{{{\color{numb}8}}}{1}
         {9}{{{\color{numb}9}}}{1}
}

\title[\texttt{pyEDITH}: exposure time calculator for HWO coronagraph]{\texttt{pyEDITH}: the coronagraphic exposure time calculator for the Habitable Worlds Observatory}

\author[Alei, Currie et al.]{
Eleonora Alei,$^{1}$\thanks{E-mail: eleonora.alei@nasa.gov}\thanks{These authors contributed equally.}\thanks{NPP Fellow}
Miles H. Currie,$^{1}$\footnotemark[2]\footnotemark[3]
Corey Spohn,$^{1}$\footnotemark[3] Christopher C. Stark$^{1}$, Aki Roberge$^{1}$, 
\newauthor  Avi M. Mandell$^{1}$, Enrico Biancalani$^{2,1,3}$, Samantha Gilbert-Janizek$^{4}$, Jacob Lustig-Yaeger$^{5}$, Sarah Steiger$^{6}$\\
$^{1}$NASA Goddard Space Flight Center, 8800 Goddard Rd, Greenbelt, 20771, MD, USA \\
$^{2}$Department of Astronomy, University of Maryland, 7901 Regents Drive, College Park, 20742, MD, USA \\
$^3$Center for Research and Exploration in Space Science and Technology II (CRESST II), Greenbelt, MD, USA\\
$^{4}$Department of Astronomy and Astrobiology Program, University of Washington, Box 351580, Seattle, Washington 98195\\
$^{5}$Johns Hopkins APL, 11100 Johns Hopkins Rd, Laurel, MD 20723, USA\\
$^{6}$ Space Telescope Science Institute, Baltimore, USA}
\date{Accepted 2026 August 26. Received 2026 August 17; in original form 2026 June 05. }

\pubyear{\the\year{}}

\begin{document}
\label{firstpage}
\pagerange{\pageref{firstpage}--\pageref{lastpage}}
\maketitle

\begin{abstract}
To support the development of next-generation missions for the search and characterization of habitable planets, high-fidelity tools for astrophysical and instrumental noise simulations are needed. In this paper, we introduce \texttt{pyEDITH}, the Python-based coronagraphic exposure time calculator built for the next recommended NASA flagship mission, the Habitable Worlds Observatory (HWO), tasked with searching for signs of habitability and life in dozens of nearby exoplanet systems. \texttt{pyEDITH} is designed to simulate wavelength-dependent exposure times and signal-to-noise ratios (S/N) for synthetic HWO direct imaging observations, considering realistic engineering specifications and user-defined target information. Its modular architecture ensures flexibility as mission requirements evolve. \texttt{pyEDITH} enables a streamlined integration with modern astronomical workflows and was designed to be used by the scientific community at all skill levels for understanding the capabilities and limitations of different HWO architectures for exoplanet analyses. The code has been validated against existing exposure time calculators and released open-source on GitHub and Zenodo, as well as made accessible through a Graphical User Interface. The \texttt{pyEDITH} package includes API documentation, tutorial notebooks, and has been used in forthcoming scientific publications.
\end{abstract}

\begin{keywords}
Software -- Algorithms -- Habitable Worlds Observatory -- Exposure Time Calculator 
\end{keywords}



\section{Introduction}
A transformative milestone in exoplanet research would be the discovery of life on other worlds. This is a complicated problem to solve: detecting an Earth-like exoplanet orbiting a Sun-like star at 10 pc distance in reflected light will require an observatory capable of reaching flux contrasts of $\sim10^{-10}$ and sub-arcsecond angular separations. This requires new instruments capable of high sensitivity in terms of flux and angular resolution.

For this reason, the Astro2020 Decadal survey recommended the development of an{ ultraviolet/optical/infrared (UV/O/IR)} Flagship Observatory to perform high-contrast direct imaging via coronagraphy, aiming to characterize at least 25 exo-Earth candidates searching for life \citep{NAP26141}. NASA responded to that recommendation by selecting the Habitable Worlds Observatory (HWO) as a future flagship mission. HWO is currently in a preliminary architecture trade phase and preparing for a Mission Concept Review by 2029 \citep{Feinberg2024,Feinberg2026}. As design trades for the HWO coronagraph are ongoing, the development of an exposure time calculator (ETC) is necessary to simulate potential HWO coronagraph designs, quantify interesting science cases, and determine the architecture that best satisfies the science requirements. 

The {HWO} community has developed a number of codes capable of exposure time calculations, such as the Altruistic Yield Optimizer \citep[AYO,][]{2014ApJ...795..122S}, the Exoplanet Open-Source Imaging Mission Simulator \citep[EXOSIMS,][]{2017ascl.soft06010S}, the Error Budget Software \citep[EBS,][]{2026arXiv260106342S}, and the \citet{Robinson2016} coronagraph model \citep[and its Python implementation described in][]{Lustig-Yaeger2019}. Those codes all serve different purposes within the HWO community and many of these ETCs were benchmarked recently by the HWO ETC Calibration Task Group \citep{2025arXiv250218556S}. The effort reflects the community's commitment to rigorous validation of simulation tools {and identifies areas of further development that these and future ETCs should explore. In addition and complementarily to these tools, we have developed \texttt{pyEDITH} (Python Exposure Direct Imaging Timer for HWO), an open-source, user-friendly ETC that is officially sponsored by the HWO Project Office, which is specifically built to interface with HWO-specific observatory specifications in a seamless way and is catered towards users at all levels of expertise.} 

\texttt{pyEDITH} is a Python package for developing the exoplanet detection and characterization capabilities of the Habitable Worlds Observatory mission. It has heritage from AYO, which is primarily used to calculate robust and fast yield calculations for HWO \citep{2014ApJ...795..122S}. Implementing this framework in Python enables easier integration with modern astronomical workflows and lowers the technical barrier for users to adopt the tool for their specific needs. \texttt{pyEDITH} was designed to allow astronomers and engineers at all career levels to study the capabilities and limitations of different HWO architectures for exoplanet detection and characterization, {as shown in several published and forthcoming scientific papers.} \texttt{pyEDITH} models noise based on the most updated HWO design specifications and through its flexible and user-friendly design, it will enable efficient HWO mission design studies and inform the development of more advanced observer planning tools when HWO launches.

In this paper, we describe \texttt{pyEDITH}'s architecture in \autoref{sec:software}. We report the results of our code validation against existing ETCs in \autoref{sec:validationetc}. We provide some use cases for the ETC as examples in  \autoref{sec:cases} and a summary and future work in \autoref{sec:summary}.

\begin{figure*}
    \centering
    \includegraphics[width=\linewidth]{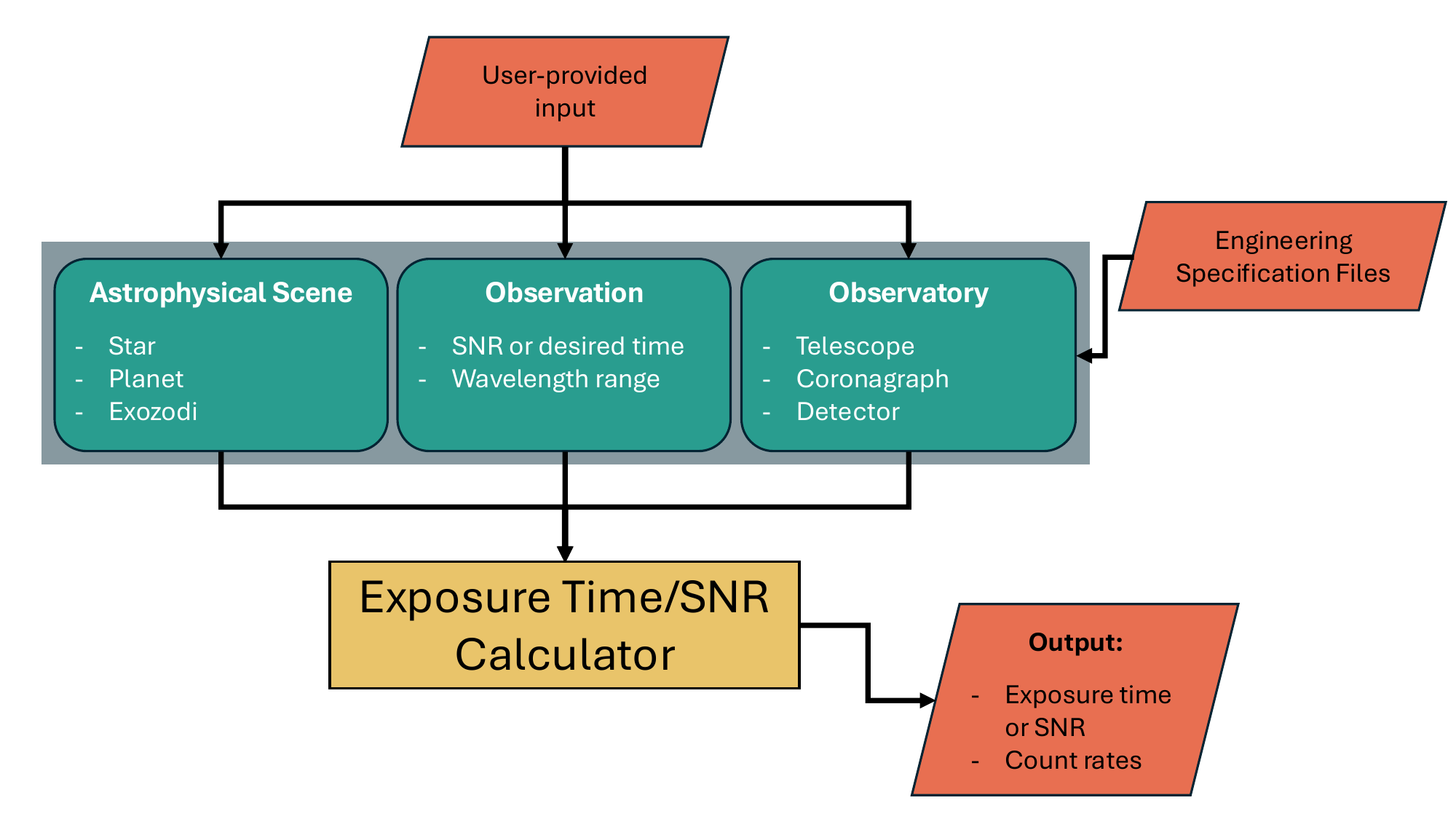}
    \caption{Simplified workflow of \texttt{pyEDITH}. Inputs can be provided via command-line configuration files or Python API. Variables are organized into three independent classes (astrophysical scene, observation parameters, observatory configuration) that communicate through a mediator (shaded region). The observatory class is subdivided into telescope, coronagraph, and detector components, typically loaded from standardized HWO files. After validation, the exposure time calculator computes observation times, signal-to-noise ratios, and noise count rates.}
    \label{fig:workflow}
\end{figure*}

\section{Software Description}\label{sec:software}

 The philosophy behind \texttt{pyEDITH}'s development is based on: 
 \begin{enumerate}
     \item being accessible by the community at large, regardless of prior knowledge of HWO architecture details;
     \item being directly connected to the HWO mission development, directly interfacing with HWO standard files providing updated estimates across different architectures;
     \item being validated with existing ETCs to ensure consistency and reliability.
 \end{enumerate}
 
\texttt{pyEDITH} was built with a modular, user-accessible design to accommodate the evolving HWO mission concept (see  \autoref{fig:workflow}).  The software performs calculations through independent but communicating classes (\autoref{sec:classes}), automatically ingesting standardized observatory specifications defined by the HWO Project Office, and then it calculates exposure times and count rates for each noise term (\autoref{sec:maths}), following the methodology used by AYO \citep{2014ApJ...795..122S, 2019JATIS...5b4009S,2025arXiv250218556S}. It is also built to provide multiple user access points (\autoref{sec:user_input}), through command line, Graphical User Interface, or any IDE. Finally, various levels of logging are implemented to customize the outputs, and unit tests are run at each release (\autoref{sec:logging}).
These features enable a diverse user base to obtain meaningful results while maintaining flexibility as the mission develops. The code is maintained on GitHub\footnote{\url{https://github.com/HabitableWorldsObservatory/pyEDITH}} and archived on Zenodo \citep{pyEDITHzenodo}.

\subsection{Classes} \label{sec:classes}

The core of \texttt{pyEDITH} is composed of three main Python classes: the astrophysical scene, the observation details, and the observatory setup. The latter is composed of three sub-classes to handle the telescope, the coronagraphic instrument, and the detector.
Variables, however they might be provided by the user (see \autoref{sec:user_input}), are fed into the relevant classes, where input is validated and checked for unit and dimension correctness. Within each class, complex variables derived from the inputs are also calculated. A ``mediator'' class is used to exchange variables from one class to the other. 

\subsubsection{Astrophysical Scene} 
This class handles the astrophysical aspects of the simulation, including stellar properties, zodiacal light, and exozodiacal light. 

The following astrophysical properties are required as input:
\begin{enumerate}
    \item the distance to the system (in parsec);
    \item the stellar radius (in $R_\odot$); 
    \item the coordinates of the star (in degrees) {assumed to be in the International Celestial Reference System (ICRS) at the J2000 epoch (neglecting stellar proper motion effects at the time of publication)};
    \item the separation of the planet from the star, provided either as angular separation in arcsec or in semi-major axis in AU ({in the semi-major axis case, the planet is assumed to be at quadrature and the separation is converted to angular separation internally}); 
    \item the amount of exozodiacal dust, expressed in units of ``zodi'', with 1 zodi being the Solar System zodiacal dust level.
    \item the stellar flux $F_*$ scaled at a distance of 10 parsec in units of $\mathrm{ph\ cm^{-2}\ s^{-1}\ nm^{-1}}$;
    \item  the planetary flux $F_\mathrm{p}$ in units of $\mathrm{ph\ cm^{-2}\ s^{-1}\ nm^{-1}}$ (or, alternatively, the planet-to-star contrast $F_\mathrm{p}/F_*$);
    \item if not included in the stellar flux array, the stellar flux in the V band scaled at a distance of 10 parsec. Otherwise, it will be taken automatically from the input stellar spectrum. 
\end{enumerate}

Stellar and planet fluxes can be provided as arrays of length equal to the input wavelength array. These will then be rebinned to the actual instrument wavelength datapoints internally.

\subsubsection{Observation} \label{sec:observation}
This class handles the observational parameters and settings for the exposure time calculator, including wavelength range and desired signal-to-noise. This class also initializes the output arrays that will be filled during the computation.

The user must provide: 
\begin{enumerate}
    \item  the wavelength array in units of $\micron$; 
    \item the desired spectral resolution $R=\frac{\lambda}{\Delta\lambda}$ for each channel (spectroscopy mode only{, see \autoref{sec:spectroscopy}}); 
    \item the lower and upper boundaries of each spectral channel (spectroscopy mode only{, see \autoref{sec:spectroscopy}});
    \item the target signal-to-noise ratio (if calculating exposure time) or exposure time (if calculating S/N); 
    \item the observing mode (\texttt{IMAGER} for photometry and \texttt{IFS} for spectroscopy).
   
\end{enumerate}

\subsubsection{Observatory}
\label{sec:eacyips}

A complete observatory system is comprised of three interdependent subsystems: the telescope, the coronagraph, and the detector. Each subsystem has its own wavelength-dependent properties that collectively determine the count rates (see \autoref{sec:maths}). 

Similarly, \texttt{pyEDITH} uses the \texttt{Observatory} class as a template for an observatory design, and represents each physical system as an abstract base class: \texttt{Telescope}, \texttt{Coronagraph}, and \texttt{Detector}. The \texttt{Observatory} abstract class acts as a container that holds one instance of each subclass (one telescope, one coronagraph, and one detector) and then calculates system-level properties. 

Since \texttt{pyEDITH} is built to interface with the latest specifications on the HWO observatory, each subclass reads in the input values from the YAML {(YAML Ain't Markup Language\footnote{\url{https://yaml.org}})} files developed for the Exploratory Analytic Cases (EACs) and hosted in the HWO Science Engineering Interface GitHub\footnote{\url{https://github.com/HWO-GOMAP-Working-Groups/Sci-Eng-Interface/tree/main}}, which must be saved locally prior to the calculation. 
Furthermore, the coronagraphic responses can be loaded from standardized Yield Input Package \citep[YIP, ][]{starkStandardizedCoronagraph} files through the \texttt{yippy}~library \citep[][see \autoref{sec:coronagraph}]{spohn_yippy_2026}.

The advanced user can, however, override any of the default variables as needed for their science case. This feature is especially useful for parameter space exploration studies.

 We have already encoded in \texttt{pyEDITH} predetermined default observatory concepts, such as the various EACs, so that the user can load the concept through a single keyword. For example, by setting this variable in a Python IDE:
 \begin{lstlisting}
 parameters['observatory_preset']='EAC5'\end{lstlisting}
 
 \texttt{pyEDITH} will load all the state-of-the-art telescope, coronagraph, and detector parameters of the EAC5 architecture as specified by the HWO YAML and coronagraph files. For more details, we point to the online documentation\footnote{\url{https://pyedith.readthedocs.io/en/latest/}}. 

\texttt{pyEDITH}~also allows the user to define a custom observatory through the special \texttt{ToyModel} class. This special class assumes wavelength-independent coronagraphic and detector responses similar in magnitude to the EAC1 concept by default, but that can be freely changed by the user to explore a larger parameter space. The Toy Model observatory has been used in recent publications to perform calculations that assume a simplified HWO design \citep[e. g.,][]{2026AJ....171..228A}.

\paragraph{Telescope}
The \texttt{Telescope} class describes the main optics and their properties. This object must be described by the following properties (either loaded from the YAML files, or user-defined):

\begin{enumerate}
    \item the primary circumscribed mirror diameter in meters; 
    \item the unobscured area of the primary mirror because of the presence of the secondary (percentage);
    \item the wavelength-dependent optical throughput;
    \item the effective throughput factor to budget for {astrophysical signal loss caused by, e. g., particulate contaminants on the primary mirror and other stray-light losses};
    \item the temperature of the optics;
    \item the estimates for the static and multiplicative overheads ($\tau_\mathrm{static}$ {[seconds]}, $\tau'_\mathrm{multi}$ {[unitless]}).
\end{enumerate}

This class calculates the collecting area of the telescope, and the thermal emission rate.

\paragraph{Coronagraph} \label{sec:coronagraph}
This class defines the coronagraph object. It requires user-specified inputs on:
\begin{enumerate}
    \item the fractional simultaneous bandwidth of the coronagraph's observing mode (e. g., 0.2 = 20\% bandwidth);
    \item the noise floor post-processing factor (PPF) used to simulate speckle subtraction (see \autoref{eq:CRbnf}).
    \end{enumerate}

In addition, the class requires 2D spatial maps of:
\begin{enumerate}
    \item stellar intensity as a function of stellar angular diameter, i. e., the coronagraph response to a point source (a star's PSF) at the center of the image;
    \item off-axis PSFs as a function of separation, i. e., the coronagraph response to an off-axis point source (a planet's PSF) at different angular separations;
    \item sky transmission maps, i. e., the response of the coronagraph to an extended source (e. g., local zodiacal light).
\end{enumerate}

These maps are part of the YIP files provided by the coronagraph developers and are generally provided with multiple solutions for varying stellar dimensions. The raw YIP files are read by the \texttt{yippy} library, which also interpolates these into maps that are valid for the specific stellar angular diameter specified by the user for the calculation.

\begin{figure*}
    \centering
    \includegraphics[width=\linewidth]{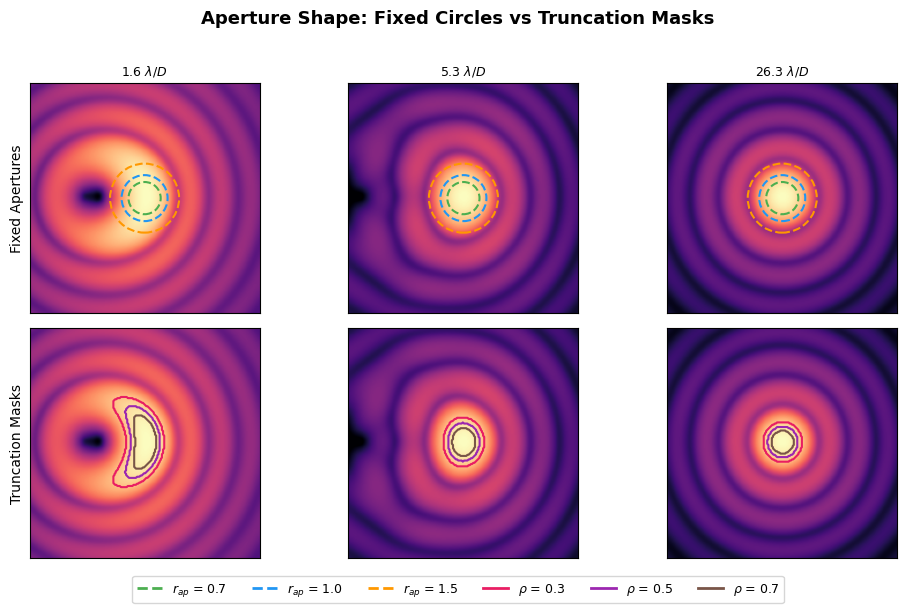}
    \caption{Comparison of the photometric aperture methods described in \autoref{sec:coronagraph} applied to three off-axis
    coronagraph PSFs of different angular separations. Each column represents an off-axis PSF loaded directly from the YIP with separation indicated by the column's header. The top row shows the fixed
    circular apertures with radii $r_\mathrm{ap}$ of 0.7, 1.0, and 1.5 $\lambda/D$ as dashed circles
    that maintain a constant shape regardless of the underlying PSF morphology.
    The bottom row shows the PSF truncation ratio masks for $\rho=0.3$, $0.5$, and
    $0.7$. The aperture mask boundary traces the threshold PSF intensity value of
    $\rho \cdot\mathrm{PSF}_\mathrm{max}$. The figure shows that the truncation ratio
    mask captures the position dependent PSF morphology near the {inner working angle (IWA)} while
    adapting to be nearly circular at large separations.}
    \label{fig:aperture}
\end{figure*}

\texttt{pyEDITH} uses two methods to define the photometric aperture, depending on the shape of the PSF itself (see \autoref{fig:aperture}), which in turn produce different values for the core throughput and core area. 

The first method uses a standard fixed circular aperture of radius $r_\mathrm{ap}$ (in $\lambda/D$), giving a core area of $\Omega=\pi r_\mathrm{ap}^2$ and calculating core throughput using a uniform aperture diameter for each planet separation. 

The second method uses the PSF ``truncation ratio'' recently added to AYO \citep{latouf_determining_2026}. The photometric aperture is defined as the set of pixels where the off-axis PSF exceeds a user-specified fraction $\rho$ of the
peak value ($\mathrm{PSF}_\mathrm{max}$). This translates into multiplying the off-axis PSF matrix by a mask defined as:
\begin{equation}
    \mathrm{mask}(x,y) =
    \begin{cases}
        1 & \text{if } \mathrm{PSF}(x{,}y) > \rho \cdot \mathrm{PSF}_\mathrm{max} \\
        0 & \text{otherwise}
    \end{cases}.
\end{equation}
This produces a core area $\Omega = \sum_\mathrm{mask} (\Delta\theta)^2$, where $(\Delta\theta)^2$ is the solid angle of one pixel, and
throughput $\eta_p = \sum_\mathrm{mask} \mathrm{PSF}(x,y)$ that both vary with separation as the PSF shape changes across the focal plane. \texttt{pyEDITH} uses the \texttt{yippy} package to compute and interpolate these quantities from the input coronagraph data.
{The user needs to provide either the \emph{photometric aperture radius} (in units of $\lambda/D$, where $\lambda$ is the wavelength of interest and $D$ is the circumscribed diameter) or the \emph{PSF truncation ratio $\rho$} (dimensionless). } If both methods are included in the input specification, \texttt{pyEDITH}
defaults to the PSF truncation ratio method and raises a warning.

Other coronagraph performance metrics such as throughput and contrast are also computed by \texttt{yippy} when the \texttt{Coronagraph} class is initialized. The discrete set of off-axis PSFs (contained in the YIP files) is processed, based on the PSF processing method specified by the user, into a set of one dimensional interpolation functions that return the performance metric as a function of separation. The user can specify that the stellar intensity map should be averaged radially, which is useful in cases when the off-axis coronagraph response is only known radially. That scenario is common for HWO coronagraphs generated with the pipeline created for the Coronagraph Design Survey \citep{belikov_coronagraph_2024}. Coronagraph models can be either hosted locally or downloaded
from \texttt{yippy}'s catalog of publicly available YIPs. They are then cached at runtime to be more easily accessed during the simulation.

 For more information, we refer to \texttt{yippy}'s documentation paper \citep[][subm. to this special issue]{spohn_yippy_2026}.

\paragraph{Detector}

This class describes the detector and its properties. The following quantities must be specified (either pulled from the HWO YAML files or custom):
\begin{enumerate}
    \item The dark current (DC) in $\mathrm{e}^-\,\mathrm{pix}^{-1}\,\mathrm{s}^{-1}$;
    \item the read noise (RN) in $\mathrm{e}^-\,\mathrm{pix}^{-1}\,\mathrm{read}^{-1}$;
    \item the read time for a single exposure ($t_\mathrm{read}$) in $s$;
    \item the clock-induced charge (CIC) in $\mathrm{e}^-\,\mathrm{pix}^{-1}\,\mathrm{ph}^{-1}$;
    \item the quantum efficiency (QE) in $\mathrm{e}^-\,\mathrm{ph}^{-1}$;
    \item the effective quantum efficiency due to degradation (dQE);
    \item the number of detector pixels per resolution element (\texttt{npix\_multiplier}), specifying the pixel binning per spectral bin (IFS mode) or spatial aperture (imaging mode).
\end{enumerate}

In the YAML files, the detector parameters are split into visible ($\lambda\leq1~\micron$) and near-infrared ($\lambda>1~\micron$), since different detectors will likely be used in the various wavelength ranges. \texttt{pyEDITH} automatically loads the relevant data based on the information provided by the user on the observation to be simulated (see \autoref{sec:observation}).

The \texttt{Detector} class calculates the physical pixel scale at a reference wavelength 500 nm based on the telescope diameter. During the exposure time calculation, the pixel scale is converted to $\lambda/D$ units for each observation wavelength, to verify that the coronagraph's photometric aperture area $\Omega$ exceeds the minimum resolvable by the detector pixel sampling, as well as to calculate the number of pixels in the photometric aperture $N_\mathrm{pix}$ for the detector noise count rate (see \autoref{eq:CRbdet}).




\subsubsection{Observatory Parameter Mediator} \label{sec:mediator}

We expect items from the three main classes (astrophysical scene, observation, and observatory) to be related to one another. For example: the coronagraph needs the telescope diameter (for arcsec to $\lambda/D$ conversions), the stellar angular diameter, the planet separation and the observation wavelength array (to interpolate the input YIP files); the detector needs the telescope diameter (to calculate the pixel scale); observation parameters are required by the observatory components (during their initialization).

Direct coupling between components (e. g., allowing a \texttt{Coronagraph} object to access the ``wavelength'' property directly from the  \texttt{Observation} instance) would create brittle code that would break any time a dependency is changed, which could happen when adding new features to accommodate a new component. This would hinder the flexibility of the code, which is one of the main requirements for \texttt{pyEDITH} to allow rapid architecture trade studies. 

\texttt{pyEDITH} implements an \texttt{ObservatoryMediator} class (shaded area in \autoref{fig:workflow}) to provide a centralized interface for cross-component parameter access. Classes request parameters from other classes through the mediator interface:

\begin{lstlisting}[language=Python]
# Inside CoronagraphYIP.load_configuration(parameters, mediator):

wavelength = mediator.get_observation_parameter("wavelength")

telescope_diameter = mediator.get_telescope_parameter("diameter")

stellar_diameter = mediator.get_scene_parameter("stellar_angular_diameter_arcsec")
\end{lstlisting}

The parameter mediator is instantiated when reading in the user input, before any calculation is performed. It stores references to all major objects and provides ``getter'' methods to return existing parameters and to enable informative error messages when parameters are not found.

\subsection{Exposure Time Calculation} \label{sec:maths}
Exposure times are calculated as \citep[following Eq. 1 of ][]{2025arXiv250218556S}:

\begin{equation}
\tau=(\mathrm{S/N})^2 \left(\frac{\mathrm{CR}_\mathrm{p}+\alpha\ \mathrm{CR}_\mathrm{b}}{\mathrm{CR}_\mathrm{p}^2 - (\mathrm{S/N})^2\ \mathrm{NF}^2}\right) \tau'_\mathrm{multi}+\tau_\mathrm{static}
\end{equation}

\noindent where S/N is the desired signal-to-noise ratio (dimensionless), $\mathrm{CR}_\mathrm{p}$, $\mathrm{CR}_\mathrm{b}$, and $\mathbf{\mathrm{NF}}$ are the photon count rates of the planet, background, and noise floor, respectively (in units of $\mathrm{\mathrm{e}^-\ s^{-1}}$). $\alpha$ parameterizes the PSF subtraction method ({values typically range between 1 and 2 depending on the PSF subtraction method, e. g., 2 for Angular Differential Imaging or 1 for a noiseless model-based PSF subtraction}), and $\tau'_\mathrm{multi}$ {(dimensionless) }and $\tau_\mathrm{static}$ {(in seconds)} are multiplicative and fixed overhead times, respectively, accounting for telescope slew/settling time and { the time needed to achieve} the required coronagraphic contrast ratio. Importantly, \texttt{pyEDITH} will also calculate S/N given a desired exposure time by inverting the equation above:

\begin{equation}
\mathrm{S/N} = \mathrm{CR}_\mathrm{p} \left[\mathrm{NF}^2 + (\mathrm{CR}_\mathrm{p} + \alpha\mathrm{CR}_\mathrm{b})\left(\frac{\tau'_\mathrm{multi}}{\tau - \tau_\mathrm{static}}\right)\right]^{-0.5}    
\end{equation}

The count rate of the planetary target is given by:
\begin{equation}
   \mathrm{CR}_\mathrm{p} = F_\mathrm{p}\ A\ \Upsilon\ T\ \Delta \lambda
   \end{equation}
   
\noindent where $F_\mathrm{p}$ is the planet flux $\left[\frac{\mathrm{photon}}{\mathrm{cm}^2 \cdot \mathrm{s} \cdot \mathrm{nm}}\right]$ at the telescope before it proceeds through the instrumentation as a function of wavelength, $A$ is the collecting area [cm$^2$], $\Upsilon$ is the fraction of light entering the coronagraph that is within the photometric core of the off-axis (planetary) PSF assuming perfectly transmitting/reflecting optics,  and $\Delta\lambda$ is the wavelength bin width [nm]. $T$ is the total throughput of the optics, calculated as the product of the combined telescope and coronagraph optical transmission, the detector quantum efficiency ($\mathrm{QE}$) and degradation quantum efficiency ($\mathrm{dQE}$), and the throughput factor that accounts for {contamination}.

The background count rate is composed of stellar leakage, local zodiacal light, exozodiacal light, observatory thermal radiation, and detector noise, and is given by:

\begin{equation}
\mathrm{CR}_\mathrm{b}= \mathrm{CR}_{\mathrm{b},*}+\mathrm{CR}_{\mathrm{b},\mathrm{zodi}}+\mathrm{CR}_{\mathrm{b},\mathrm{exozodi}}+\mathrm{CR}_{\mathrm{b},\mathrm{thermal}}+\mathrm{CR}_{\mathrm{b},\mathrm{detector}}
\end{equation}

Coronagraphs cannot block all host star light, and so the stellar leakage term is given by:

\begin{equation}
\mathrm{CR}_{\mathrm{b},*} = F_{*} \ \frac{I_{*}}{\theta^2} \Omega\ A\ T\ \Delta \lambda
\end{equation}

\noindent where $F_*$ is the stellar flux $\left[\frac{\mathrm{photon}}{\mathrm{cm}^2 \cdot \mathrm{s} \cdot \mathrm{nm}}\right]$ as a function of wavelength,{ $\frac{I_{*}}{\theta^2}$ is the spatially dependent leaked stellar count rate per unit solid angle exiting the instrument normalized to the starlight entering the instrument \citep[following the notation used in][]{2019JATIS...5b4009S},} and $\Omega$ is the area of the photometric aperture in steradians.
{$I_{*}$ is the stellar intensity map provided by the YIP files read by the \texttt{Coronagraph} class. In the ``Toy Model'' coronagraph case, we approximate the value of $\frac{I_{*}}{\theta^2}$ with $\zeta\,\mathrm{PSF}_\mathrm{peak}$, where $\zeta$ is the uniform level of suppressed starlight relative to the peak of the PSF \citep[following][]{2014ApJ...795..122S}. 
  }

The count rate of the solar system zodiacal dust, assumed to be a gray scatterer, is given by:
\begin{equation}
\mathrm{CR}_{\mathrm{b},\mathrm{zodi}}= F_0\ 10^{-0.4z } \Omega\ A\ T\ T_\mathrm{sky} (x, y)\ \Delta \lambda
\end{equation}

\noindent where $F_0$ is the zero-point flux $\left[\frac{\mathrm{photon}}{\mathrm{cm}^2 \cdot \mathrm{s} \cdot \mathrm{nm}}\right]$ as a function of wavelength{; $T_{sky}$ is the sky transmission map provided by the YIP file,} $z$ is the V-band surface brightness of the zodi $\left[\mathrm{mag}/\mathrm{arcsec}^2\right]$ scaled by the zodi optical depth integrated along the target line of sight, {and $(x, y)$ the pixel coordinates of the planet in the image}. {To do this, we interpolate the values in Table 17 of \citet{1998A&AS..127....1L}, expressed in terms of ecliptic latitude $\beta$ and longitude relative to the Sun $\Delta\lambda_\odot=\lambda-\lambda_\odot$. We assume to observe at a fixed  $\Delta\lambda_\odot=135^\circ$, where the zodiacal light is minimum and only slowly varying with longitude, and we interpolate on $\beta$ based on the user-provided coordinates (converted to the ecliptic reference frame internally).}

The count rate of dust in exoplanet systems is analogous to the zodiacal light, and is given by:

\begin{equation}
\mathrm{CR}_{\mathrm{b},\mathrm{exozodi}}= F_0\ n 10^{-0.4z' (x, y)} \Omega\ A\ T\ T_{sky} (x, y)\ \Delta \lambda
\end{equation}

\noindent where $z'$ is the surface brightness in the V band of the exozodi, {normalized to be }$22\ \mathrm{mag}/\mathrm{arcsec}^2$ {at 1 AU for a Solar twin. The exozodi disk is assumed to be decreasing in brightness with $r^{-2}$, where $r$ is the radial distance, following the approach used in \citet{2014ApJ...795..122S}}. $n$ is the exozodi multiplier, which controls the density of exozodiacal dust in a system as a multiple of zodiacal dust.

The thermal emission of the observatory is given by:
\begin{equation}
\mathrm{CR}_{\mathrm{b},\mathrm{thermal}}= \frac{B_\lambda}{E_\mathrm{photon}} \ \varepsilon_\mathrm{warm}\ T_\mathrm{cold}\ \mathrm{QE'}\ \Omega\ A\ \Delta \lambda
\end{equation}

\noindent where $B_\lambda$ is the spectral radiance per wavelength (according to Planck's law and depending on temperature); $E_\mathrm{photon}$ is the energy of the photon; $\varepsilon_\mathrm{warm}$ is the effective emissivity of all warm optics; $T_\mathrm{cold}$ is the transmission/reflectivity of all cold optics; $\mathrm{QE'}$ is the {detector's effective quantum efficiency (raw QE multiplied by the QE degradation factor dQE)}.

Noise from the detector is given by:

\begin{equation}\mathrm{CR}_{\mathrm{b},\mathrm{detector}}= N_\mathrm{pix} \left(\mathrm{DC}+\frac{\mathrm{RN}^2}{t_\mathrm{read}}+\frac{\mathrm{CIC}}{t_\mathrm{count}}\right)
\label{eq:CRbdet}
\end{equation}

\noindent where $N_\mathrm{pix}$ is the number of detector pixels in the photometric aperture; $\mathrm{DC}$ is the dark current (i. e., thermally generated electrons) [$\mathrm{e}^-$/pix/s],{ $\mathrm{RN}^2$ is the read noise variance (introduced during the detector readout process) [$\mathrm{e}^-$/pix/read]}, $t_\mathrm{read}$ is the read time (i. e., the time necessary to read out the entire detector array) [s], $\mathrm{CIC}$ is the clock-induced-charge (i. e., spurious electrons generated during pixel clocking operations) [$\mathrm{e}^-/\mathrm{pix}/\mathrm{photon}$]; and $t_\mathrm{count}$ is the average time to detect one photon per pixel (which determines how often clock-induced-charge events occur) [s].

The noise floor count rate simulates imperfect coronagraphic speckle subtraction and is given by:
\begin{equation}
\mathrm{NF}= \mathrm{CR}_{\mathrm{b},*} / \mathrm{PPF} \label{eq:CRbnf}
\end{equation}

\noindent where $\mathrm{PPF}$ is an assumed post-processing factor, nominally 30 for HWO, assuming $10^{-10}$ raw contrast is achieved. {The use of a post-processing factor to obtain a noise floor level proportional to the raw contrast is well-established in the literature \citep{Nemati2014,Garrett2017b,nemati2023analytical}.}


\vspace{0.5em}
Once the calculation is done, \texttt{pyEDITH}~ returns the desired output (exposure time or signal-to-noise ratio), as well as every count rate and some ancillary variables that could be used for validation. We use these variables to validate \texttt{pyEDITH}~with existing ETCs (see \autoref{sec:validationetc}).

The calculation works seamlessly for both scalar objects and arrays. For this reason, the same code can be used for both the coronagraphic imaging mode and spectroscopy mode.

\subsection{User Interfaces}\label{sec:user_input}

{Users can provide inputs} either through a text input file which can be read by command line, or by providing a Python dictionary when calling the Python API. This allows a variety of users to access \texttt{pyEDITH}, without prior knowledge of Python necessary.

For the command-line case, the user can run three functions depending on the calculation to be executed.
\begin{lstlisting}[language=sh]
pyedith [-h] [-v] [-q] {etc,snr,etc2snr} --edith INPUT_FILE [--time TIME]
\end{lstlisting}

The \texttt{etc} function calculates the exposure time for specific wavelength(s); the \texttt{snr} function calculates the signal-to-noise ratio given a specific observing time (to be specified in seconds through the optional argument \texttt{time}); the \texttt{etc2snr} function allows the user to calculate the exposure time for a primary wavelength and to assume this quantity to calculate the signal-to-noise ratio for a secondary wavelength. All these functions require an input file specified under the argument \texttt{edith}. See \autoref{fig:inputfile} for an example input file.

Other relevant flags when running \texttt{pyEDITH} from command line are: \texttt{-h} for the help statements, \texttt{-v/-vv} for increased levels of verbosity (\texttt{-v} for information, \texttt{-vv} for debug logs with detailed output), \texttt{-q} for a quiet mode that does not show warnings but only errors (see \autoref{sec:logging} for details).

When running using the Python API, the user must provide a Python dictionary of parameters (\texttt{params} hereafter) and then call specific functions for the calculation of exposure time or signal-to-noise ratio. For the latter, the exposure time must be provided as an astropy \texttt{Quantity} of time \citep{astropy2013, astropy2018, astropy2022}, which allows the user to specify the value in any time unit (e. g., seconds, hours). The dictionary is parsed by \texttt{pyEDITH} to ensure consistency for the required format.
Both functions return a \texttt{validation\_output} quantity, a dictionary that contains all relevant variables with units, useful for debugging. For example, given a dictionary of parameters, exposure time or S/N can be calculated as follows:

\begin{lstlisting}[language=Python]
from pyEDITH import parse_input, calculate_texp, calculate_snr
from astropy import units as u

parsed_parameters= parse_input.parse_parameters(params)
texp, validation_output = calculate_texp(parsed_parameters)
print(f"Calculated exposure time: {texp.to(u.hr)}")

texp = 3*u.hr
snr, validation_output = calculate_snr(parsed_parameters, texp)
print(f"Calculated S/N: {snr}")

\end{lstlisting}

\texttt{pyEDITH} can also be accessed via the Graphical User Interface {(GUI)} of HWO\footnote{Imaging mode: \url{https://hwo.stsci.edu/coron_imaging}; spectroscopy mode: \url{https://hwo.stsci.edu/coron_spec}}. At the time of publication, the GUI allows users to pick a template for star and planet or load custom datasets, and to calculate exposure times varying observatory specifications. The GUI is maintained by the Space Telescope Science Institute and not part of the base \texttt{pyEDITH} repository.

\subsection{Logging and Unit Testing} \label{sec:logging}

\texttt{pyEDITH} uses Python's standard \texttt{logging} module to provide runtime diagnostics. Log messages are divided into:
\begin{enumerate}
\item \texttt{info} messages for configuration and methodology;
\item \texttt{warning} messages for non-fatal issues that require attention;
\item \texttt{error} messages for calculation failures that cause invalid results;
\item \texttt{debug} messages that {print} variables for validation.
\end{enumerate}

During runtime, the code enforces a check on the existence and the dimensions of each variable via \texttt{astropy.units} type checking. 

The code is fully tested with unit tests that run at every commit to the GitHub repository through continuous integration. Tests use mock objects to isolate external dependencies. The dimensionality of each variable is also checked in the unit tests, together with a numerical accuracy validated to relative tolerances of $\le0.01\%$.

\section{Validation}\label{sec:validationetc}

\begin{figure*}
    \includegraphics[width=\linewidth]{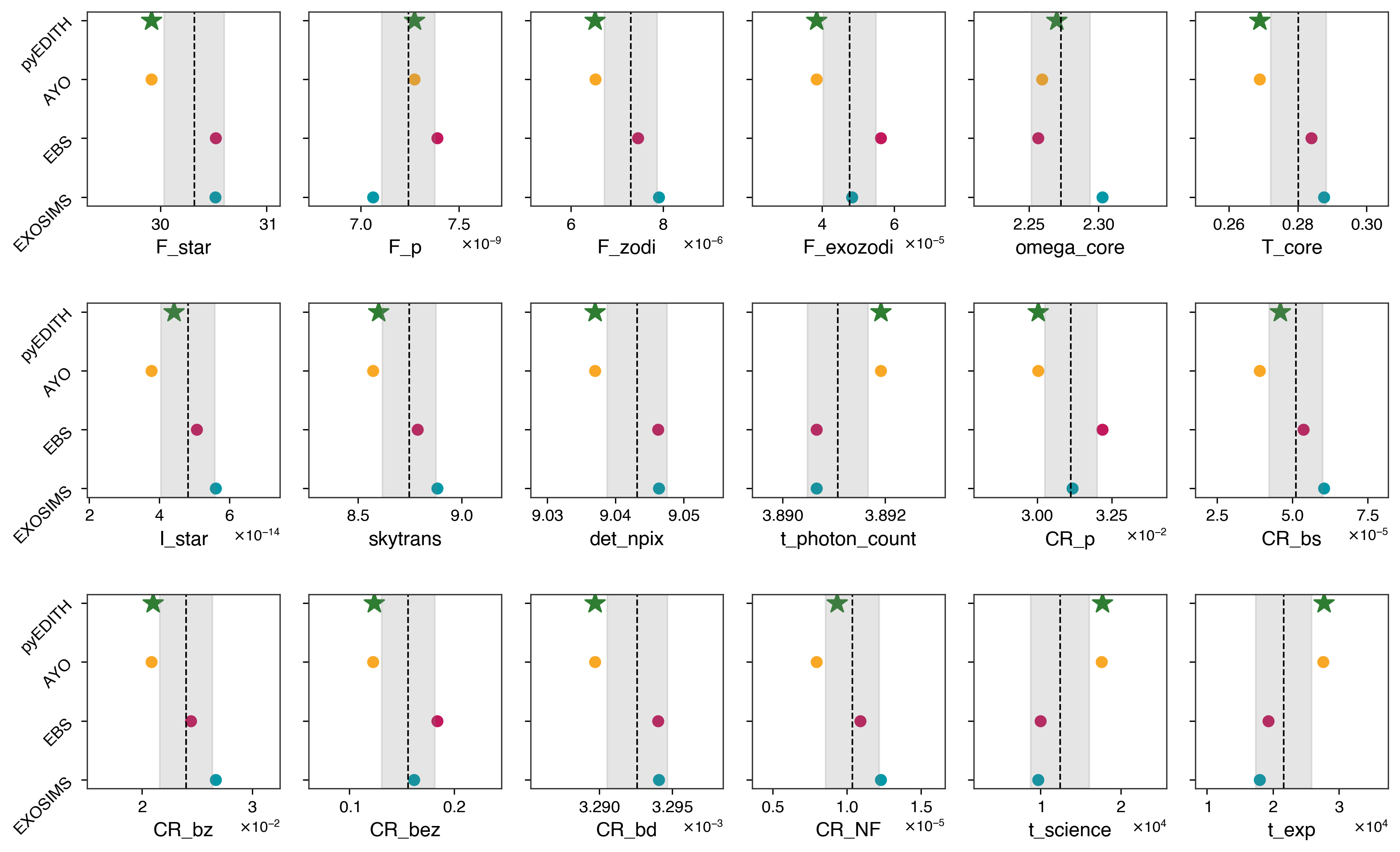}
    \caption{Comparison between \texttt{pyEDITH} and other ETCs \citep[from][]{2025arXiv250218556S}. Markers show: results from \texttt{pyEDITH} (green stars); AYO (yellow circles); EBS (red circles); EXOSIMS (cyan circles). As a dashed black line, the mean value of the estimates obtained by AYO, EBS, and EXOSIMS; as a shaded gray area, the 1-$\sigma$ envelope of these estimates.}\label{fig:validation}
\end{figure*}

\citet{2025arXiv250218556S} recently compared existing codes that internally use exposure time calculators to perform calculations for HWO (AYO, EXOSIMS, and EBS). We followed the same validation process with \texttt{pyEDITH}, confirming a general agreement between our code and the others. As part of the \texttt{pyEDITH}~ testing pipeline, we reproduced the validation scenarios compared in \citet{2025arXiv250218556S} and found a reasonable agreement. We show in \autoref{fig:validation} the results of one of the validation cases performed in  \citet{2025arXiv250218556S}. Figures for other cases can be found in the online documentation\footnote{\url{https://pyedith.readthedocs.io/en/latest/validation.html}}.

Since \texttt{pyEDITH}~stems from AYO, we expect to find the smallest differences between AYO and \texttt{pyEDITH}. However, some differences appear in the calculation of the zero-point flux $F_0$ between \texttt{pyEDITH} and AYO, given fundamental differences in the interpolation algorithm between Python and IDL (the language AYO is written in). All variables derived from the zero-point flux (the stellar leakage, the exozodi, the noise floor count rates, as well as the exposure time) appear therefore to be different up to 10\%, a disagreement that was still considered acceptable in the general ETC comparison performed by \citet{2025arXiv250218556S}.

Some coronagraph-related variables (stellar intensity \texttt{I\_star}, core area \texttt{omega\_core}) are also different since \texttt{pyEDITH}~relies on \texttt{yippy}~for the calculation of these variables, which is marginally different from the original AYO calculation. The differences are most pronounced in cases where the YIP only includes the coronagraph's off-axis PSFs along one dimension (e. g., every PSF is along the x-axis). In that scenario \texttt{yippy} calculates and uses the radial average of the stellar PSF, whereas AYO uses the full 2D stellar PSF when calculating exposure times.

In the cases where the values produced by \texttt{pyEDITH} do not match those produced by AYO, \texttt{pyEDITH}'s estimates still lie within the 1-$\sigma$ error of the mean of all previous values. This shows robust agreement with existing codes.

\section{Use Cases} \label{sec:cases}

To showcase the versatility of \texttt{pyEDITH} for mission design and observation planning, we present a series of use cases from the online tutorials and from published papers that used this ETC. These examples illustrate how the tool can be used to perform trade studies, analyze noise sources, and produce simulated observations, all while maintaining consistency with HWO design specifications.

\subsection{Noise Budget Analysis}

\begin{figure*}
    \centering
    \includegraphics[width=\linewidth]{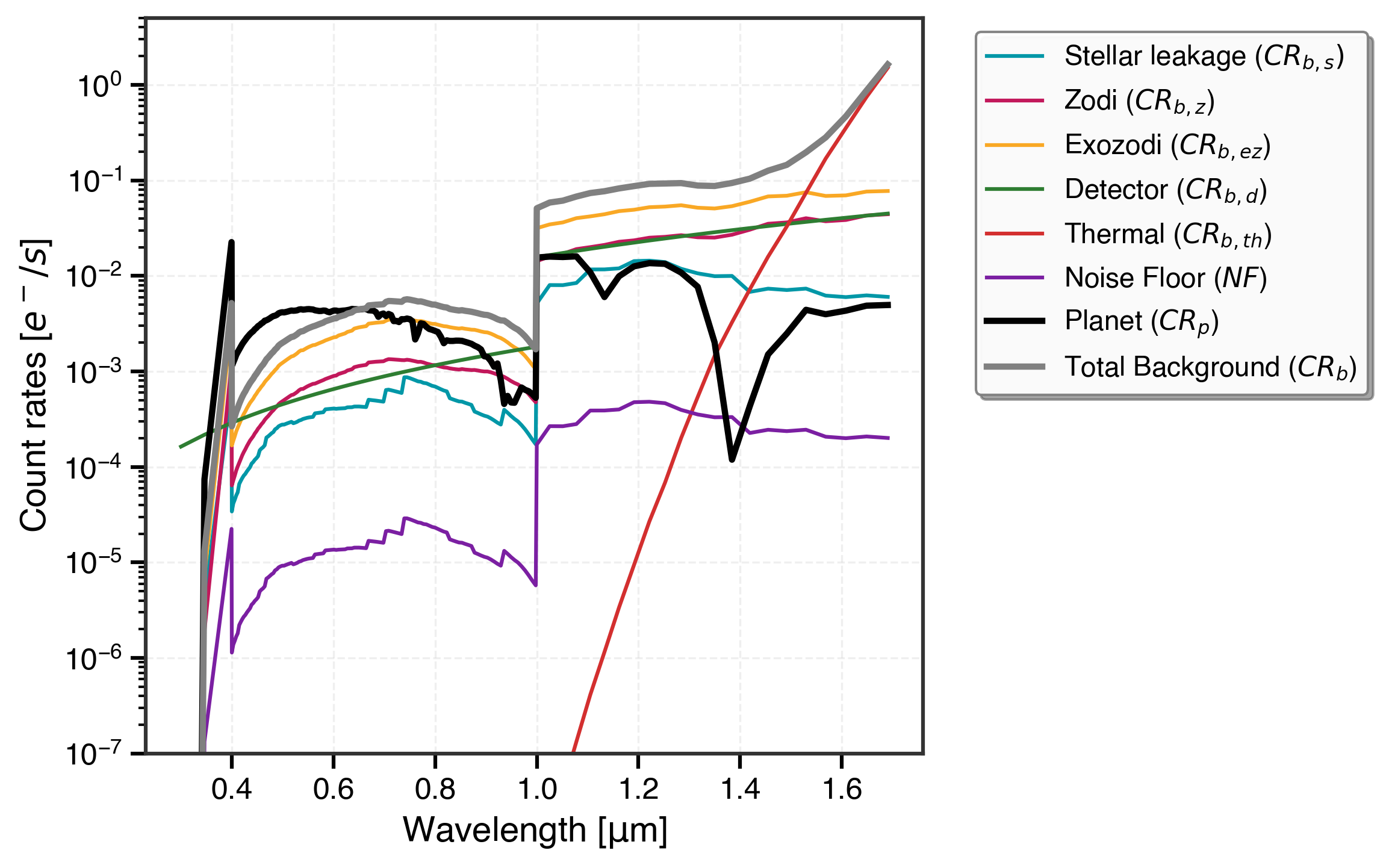}
    \caption{Count rate budget for a simulated Earth-twin { (1 $R_\oplus$, 1 $M_\oplus$) at quadrature }around a Sun-like star {(blackbody of temperature 5880 K with radius 1 $R_\odot$, coordinates [176.63,-40.50] degrees)} at 10 pc distance, and {exozodi level equal to 3 zodis}, obtained using \texttt{pyEDITH}. {The simulation assumed the EAC1 concept in its default configuration, with 7.2 m circumscribed primary mirror size; 270 K temperature of the primary mirror; and using the EAC1 off-axis amplitude-apodized vortex coronagraph \citep{10.1117/1.JATIS.12.4.041011}. } Individual noise components (stellar leakage, zodiacal/exozodiacal light, thermal emission, detector noise, noise floor) are plotted alongside the planet signal $\mathrm{CR}_\mathrm{p}$ and the total background noise $\mathrm{CR}_\mathrm{b}$. }
    \label{fig:countrates}
\end{figure*}

Understanding the dominant noise sources is critical for the ongoing coronagraphic studies for HWO. In \autoref{fig:countrates}, we plot the photon count rates for an Earth-twin as a function of wavelength. We report all the terms that are internally calculated by \texttt{pyEDITH} (see \autoref{sec:maths}) and we visualize them simultaneously. This way, users can quickly identify which components dominate the noise budget as a function of wavelength. This is an important diagnostic when optimizing the observatory parameters.

\subsection{Spectroscopic Simulations}\label{sec:spectroscopy}

\begin{figure*}
    \centering
    \includegraphics[width=0.8\linewidth]{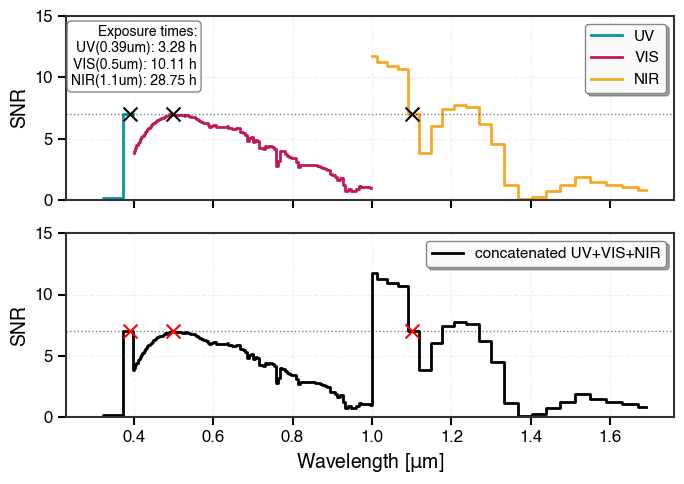}
    \caption{S/N calculation across different channels {for a simulated Earth-twin (1 $R_\oplus$, 1 $M_\oplus$) at quadrature around a Sun-like star {(blackbody of temperature 5880 K with radius 1 $R_\odot$, coordinates [176.63,-40.50] degrees)} at 10 pc distance, and exozodi level equal to 3 zodis, obtained using \texttt{pyEDITH}. The simulation assumed the EAC1 concept in its default configuration, with 7.2 m circumscribed primary mirror size; 270 K temperature of the primary mirror; and using the EAC1 off-axis amplitude-apodized vortex coronagraph \citep{10.1117/1.JATIS.12.4.041011}}. The top panel illustrates independent S/N calculations for the ultraviolet (UV), visible (VIS), and near-infrared (NIR) channels, with the required exposure times to achieve S/N=7 at the specified reference wavelengths (labeled as X markers) shown in the top left corner. The bottom panel shows the concatenated S/N result. }
    \label{fig:snr}
\end{figure*}

One of the primary requirements for HWO is to characterize exoplanet atmospheres via direct imaging spectroscopy. The spectroscopy mode of \texttt{pyEDITH} calculates exposure time and S/N as a function of wavelength given user-defined models of the host star and exoplanet reflectance spectra. The user can define spectral channels and their corresponding resolutions, enabling maximum flexibility for spectroscopic instrumentation trade studies. We show in \autoref{fig:snr} the S/N calculated in the UV, visible, and NIR channels based on the exposure time that is necessary to reach the threshold value of S/N=7 at a specific reference wavelength within each channel. This feature allows mission planners to evaluate the feasibility of multi-bandpass characterization strategies.

\begin{figure*}
    \centering
    \includegraphics[width=0.8\linewidth]{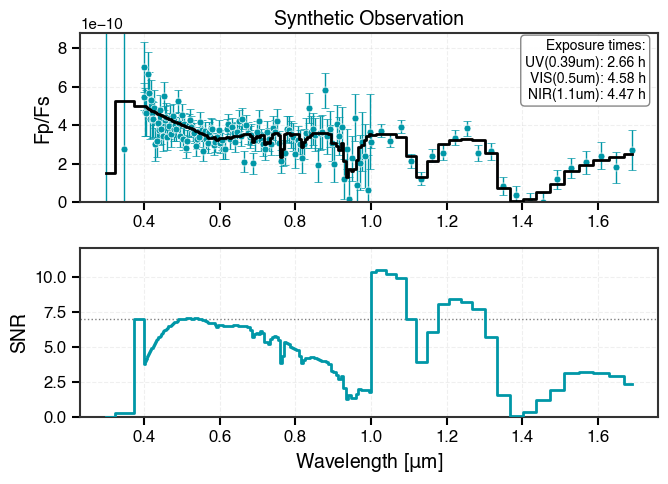}
    \caption{Example of \texttt{pyEDITH}-synthesized HWO data (cyan) of an Earth-like exoplanet (upper) and the corresponding S/N as a function of wavelength (lower). {For a simulated Earth-twin (1 $R_\oplus$, 1 $M_\oplus$) at quadrature around a Sun-like star {(blackbody of temperature 5880 K with radius 1 $R_\odot$, coordinates [176.63,-40.50] degrees)} at 10 pc distance, and exozodi level equal to 3 zodis. The simulation assumed the EAC1 concept in its default configuration, with 7.2 m circumscribed primary mirror size; 270 K temperature of the primary mirror; and using the EAC1 off-axis amplitude-apodized vortex coronagraph \citep{10.1117/1.JATIS.12.4.041011}}. Like in \autoref{fig:snr}, the S/N is calculated separately for the UV, VIS, and NIR channels by integrating for the exposure time necessary to reach S/N=7 in specific reference wavelengths (in parentheses).}
    \label{fig:spec}
\end{figure*}

Importantly, \texttt{pyEDITH} can synthesize noisy exoplanet observational data to use in data analysis simulations (see \autoref{fig:spec}).  Data can be automatically randomized within \texttt{pyEDITH} if the user desires, thus simulating a realistic observation.

\subsection{Architectural Trade Studies}

Beyond individual observations, \texttt{pyEDITH} is designed to perform rapid parameter space exploration to support the ongoing trade studies for the definition of HWO's final design. Since the tool is modular and can be easily customized by overriding default observatory parameters, it can quickly perform parameter studies for various concepts.

\begin{figure*}
    \centering
    \begin{subfigure}{0.48\textwidth}
        \centering
        \includegraphics[width=\textwidth]{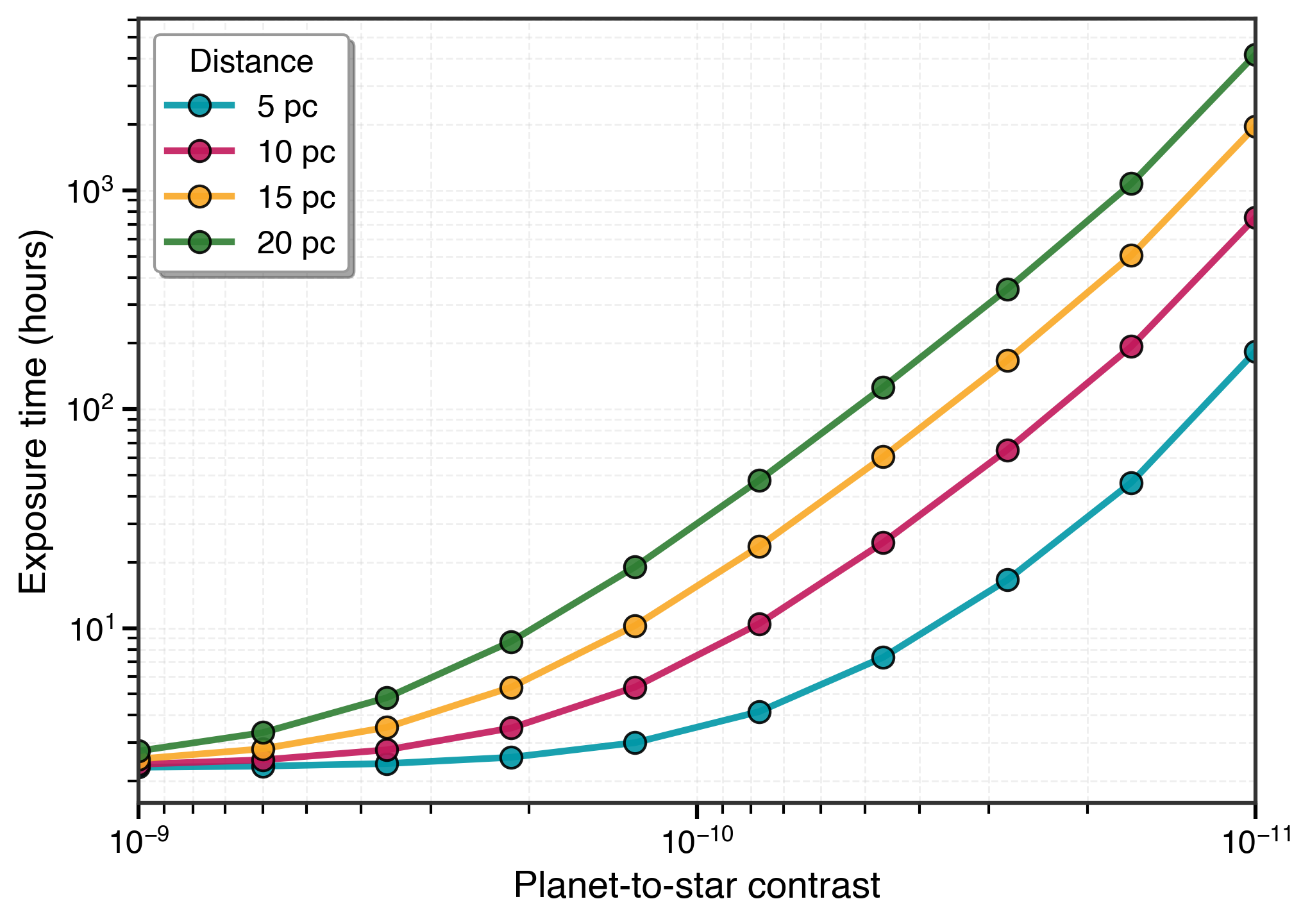}

    \end{subfigure}
    \hfill
    \begin{subfigure}{0.48\textwidth}
        \centering
        \includegraphics[width=\textwidth]{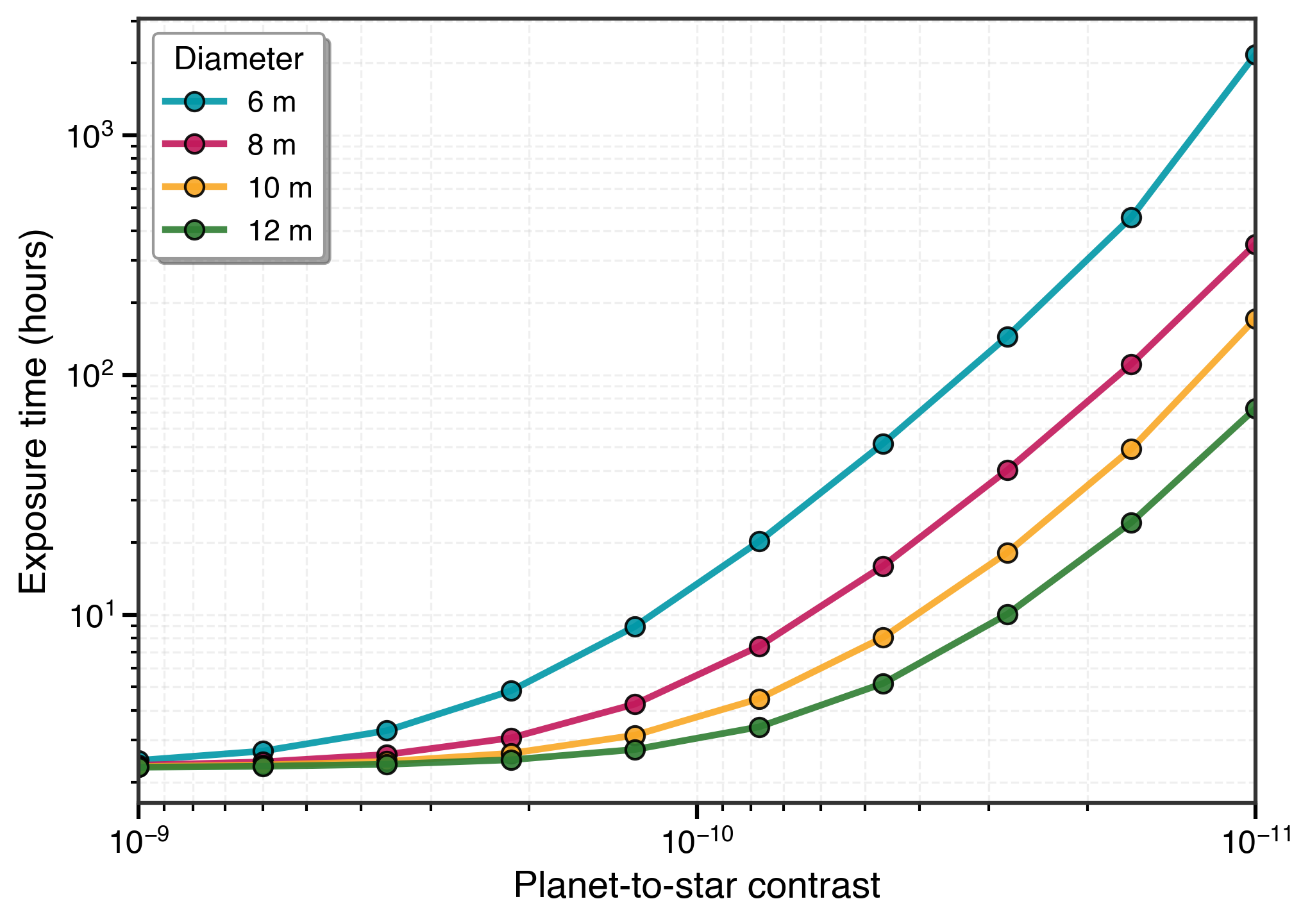}

    \end{subfigure}
    \caption{{Example plots useful for architecture trades. \emph{Left}: Impact of system distance on the exposure time required to reach S/N=7 at 500 nm in broadband photometry for an Earth-twin (1 $R_\oplus$, 1 $M_\oplus$)  at quadrature around a Sun-like star (blackbody of temperature 5880 K with radius 1 $R_\odot$, coordinates [176.63,-40.50] degrees) and exozodi level equal to 3 zodis, assuming the EAC1 concept in its default configuration, with 7.2 m circumscribed primary mirror size; 270 K temperature of the primary mirror; and using the EAC1 off-axis amplitude-apodized vortex coronagraph \citep{10.1117/1.JATIS.12.4.041011}. Exposure time is plotted as a function of planet-to-star contrast and for varying system distances (5 to 20 pc). For the same contrast level, exposure time increases with the distance to the stellar system. \emph{Right}: Effect of aperture size on the exposure time required to reach S/N=7 at 500 nm in broadband photometry for an Earth-twin (1 $R_\oplus$, 1 $M_\oplus$)  at quadrature around a Sun-like star (blackbody of temperature 5880 K with radius 1 $R_\odot$, coordinates [176.63,-40.50] degrees) at 10 pc distance, and exozodi level equal to 3 zodis. We assumed the EAC1 concept in its default configuration with 270 K temperature of the primary mirror; and using the EAC1 off axis amplitude-apodized vortex coronagraph (see above for reference), but changing the diameter of the primary mirror. Exposure time is plotted as a function of planet-to-star contrast and for varying telescope diameters (6 to 12 meters). Larger apertures reduce the overall exposure time and are more sensitive to smaller contrasts.}}\label{fig:trades}
\end{figure*}

{In \autoref{fig:trades} we show two examples of possible trades (exposure time as a function of planetary contrast, assuming a default EAC1 observatory with a broadband photometry observation of an Earth-twin at 500 nm, and varying either the telescope diameter or the distance to the system). Such trades can be easily performed with \texttt{pyEDITH} and could provide relevant insight on the coronagraph performance as both astronomical and observatory parameters vary.}

\subsection{Instrumentation Development}

\texttt{pyEDITH}'s modular utility functions can also be leveraged by custom software tools for instrument trade studies.
For example, \citet{Biancalani2025} utilized \texttt{pyEDITH}'s Gaussian-convolution spectral re-gridding function (\texttt{regrid\_spec\_gaussconv}) to develop trade-off simulations for O$_2$ detection in an exo-Earth, observed in reflected starlight around a Sun-like star with HWO/EAC1. Figure~6 of \citet{Biancalani2025} demonstrates how spectral resolving power trades against detector noise characteristics (dark current and read noise), informing the design of spectrographs capable of operating at different spectral resolutions.

\subsection{Observation Planning}

\begin{figure*}
    \centering
    \includegraphics[width=0.8\linewidth]{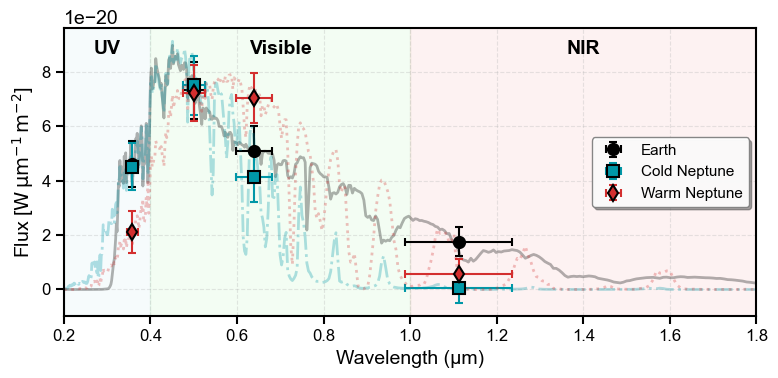}
    \caption{Simulated photometric bandpasses to qualitatively characterize different spectra (Earth {at quadrature at 1 AU} in black, Cold Neptune {at phase 7$^\circ$, inclination 97$^{\circ}$ and distance} 9 AU in cyan, Warm Neptune  {at phase 155$^\circ$, inclination 65$^\circ$ and distance} 2.25 AU in red). Horizontal error bars show the width of the bandpass; vertical error bars represent the associated noise. The selected bandpasses are the optimal bandpasses for class differentiation obtained in \citet{2026AJ....171..228A}. {These spectra assume a Toy Model HWO concept (primary mirror:  7.87 m circumscribed diameter, 290 K temperature, 87.9\% unobscured area) and a Toy Model coronagraph, considering a Sun-like host star at 10 pc distance, following the paper's prescriptions. We refer to Table 2 in \citet{2026AJ....171..228A} for details.}
 }
    \label{fig:imaging}
\end{figure*}

In broadband photometry mode, \texttt{pyEDITH} can calculate the exposure time needed to reach the S/N required for initial exoplanet detection surveys in any bandpass for user-defined exoplanet systems. This has been used in \citet{2026AJ....171..228A} to evaluate different filter combinations that, when used in parallel, would provide color information capable of distinguishing different planetary classes, thus optimizing the survey strategy. We show in \autoref{fig:imaging} simulated HWO broadband photometry parallel measurements that are promising for differentiating an Earth twin from two Neptune planets, according to the results in \citet{2026AJ....171..228A}. Practically, this result has been obtained by using the \texttt{etc2snr} function in \texttt{pyEDITH}, first determining the exposure time necessary to reach S/N=7 at 500 nm, and then using that time to evaluate the noise on other bandpasses.

\subsection{Spectral Characterization Studies}

\begin{figure*}
    \centering
    \includegraphics[width=\linewidth]{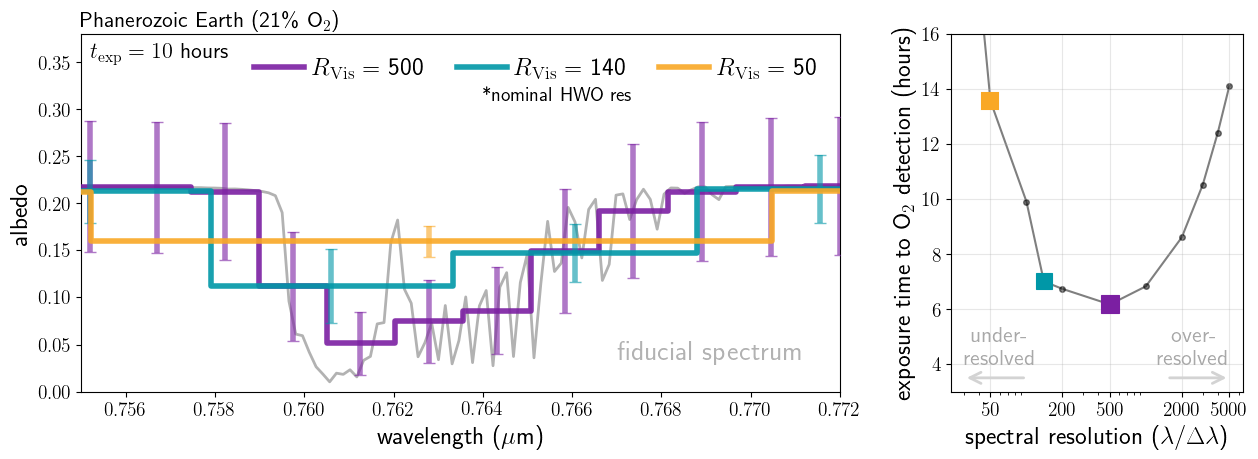}
    \caption{Impact of spectral resolution on the exposure time and detectability of spectral features. \emph{Left}: O$_2$ band at 0.76 micron for a Phanerozoic Earth model {at quadrature at 10 pc distance orbiting a Sun-like star,} for varying resolutions assuming a constant exposure time of 10 hours.  The lowest resolution ($R_\mathrm{Vis}=50$, yellow) has the fewest data points and the smallest error bar, but the depth of the line is more easily confused with the continuum. Increasing resolution ($R_\mathrm{Vis}=140$, green, and $R_\mathrm{Vis}=500$, purple) translates into adding more spectral data points to map the line depth more accurately, but also increases the noise envelope for each point. \emph{Right}: Exposure time required to correctly detect O$_2$ for each of the spectral resolutions in the left panel. There is a sweet spot between under-resolving and over-resolving a spectral feature which minimizes the exposure time for detection. {These calculations assumed an EAC1-like telescope (primary mirror:  7.22 m circumscribed diameter, 270 K temperature) assuming a LUVOIR-B-like coronagraph, varying resolutions and dark current parameters. We refer to Table 2 of  \citet{2026arXiv260426925G} for details.}}
    \label{fig:sams}
\end{figure*}

\texttt{pyEDITH} is a key tool to be used in spectral characterization studies and retrievals performed to simulate the expected performance of HWO. \citet{2026arXiv260426925G} used the ETC to simulate HWO EAC1 noise levels in the visible and near-infrared range at varying assumptions of spectral resolutions (R=20 to 5000) and dark current values, for five atmospheric scenarios spanning Earth's evolution. The noisy spectra generated with \texttt{pyEDITH} were then studied to assess the impact of under- or oversampling of spectral lines (determined by the resolution, see \autoref{fig:sams}) and detector performance (determined by dark current) when constraining relevant biosignature gases. Their results validate HWO's baseline resolution choices while identifying critical trade-offs for instrument design.

In \citet{2026arXiv260714329C}, \texttt{pyEDITH} was employed to generate noisy synthetic spectra across various visible and infrared channels \citep[see Figure 1 of][]{2026arXiv260714329C} under diverse exozodi scenarios and density levels. By performing atmospheric retrievals on these data, the authors demonstrated that exozodi can mask absorption features or mimic atmospheric properties, leading to significant biases in atmospheric composition estimates. This study was instrumental in refining the exozodi treatment within \texttt{pyEDITH}, specifically driving the implementation of diverse dust populations (gray, red, and blue dust) and processing treatments. These enhancements allowed for the identification of critical correlations between dust color, post-processing, and the morphology of spectral lines, providing a framework to assess the impact of exozodiacal light on future reflected-light observations with HWO.

\section{Summary and Future Work} \label{sec:summary}

In this paper, we introduced \texttt{pyEDITH}, a Python-based exposure time calculator for the HWO coronagraph that has been developed to support mission design and science trade studies. \texttt{pyEDITH} calculates noise budgets and exposure times by ingesting standardized data from the HWO architectural concepts developed by the HWO Project Office. We described the software implementation, demonstrated its utility through scientific use cases, and validated it against existing tools, confirming its maturity in this first release. 

As HWO matures towards its Mission Concept Review, \texttt{pyEDITH} will continue to evolve. Our development roadmap prioritizes the following objectives:
\begin{enumerate}
    \item establishing common repositories for alternative coronagraph and detector specifications, in coordination with the HWO Coronagraph Coordination Group and instrument developers;
    \item automating the ingestion of stellar and planetary parameters from existing astronomical databases;
    \item refining the noise floor treatment by incorporating higher-fidelity speckle distribution maps as they become available;
    \item validation against emerging simulation tools (e. g., \texttt{coronagraphoto}\footnote{\url{https://github.com/CoreySpohn/coronagraphoto}}) and real observations such as Roman CGI results;
    \item enhancing diagnostics and logging to better alert users when input overrides deviate from validated HWO design envelopes;
    \item hosting periodic community tutorials to encourage code adoption and gather feedback.

\end{enumerate}

Following our software development philosophy of open, reproducible software, \texttt{pyEDITH} aims to lower the technical barriers for community engagement, broadening participation in the design and scientific planning of the next flagship observatory.

\section*{Acknowledgements}

We thank the anonymous reviewers for their helpful feedback. We acknowledge helpful feedback from Adric Riedel, Andrew Myers, and Jason Tumlinson and the broader Habitable Worlds Observatory community. We thank the beta testers Amber Young, Connor Vancil, Giada Arney, Joshua Krissansen-Totton, Laurent Pueyo, Natasha Latouf, Rahul Arora, Katherine Costello, and Tyler Robinson. The work of E. A., M. H. C., and C. S. was supported by appointments to the NASA Postdoctoral Program at the NASA Goddard Space Flight Center, administered by Oak Ridge Associated Universities under contract with NASA (ORAU-80HQTR21CA005).

During the preparation of this work the authors used Large Language Models developed at NASA in order to produce code documentation, refactor code, and produce unit tests. After using this tool/service, the authors reviewed and edited the content as needed and take full responsibility for the content of the publication.

\section*{Data Availability}

 The \texttt{pyEDITH} source code is publicly available on GitHub\footnote{\url{https://github.com/HabitableWorldsObservatory/pyEDITH}}. The code has been archived on {Zenodo \citep{pyEDITHzenodo}} with the DOI 10.5281/zenodo.17917471. Users may also access the tool through the HWO community-hosted Graphical User Interface\footnote{\url{https://hwo.stsci.edu/coron_imaging} and \url{https://hwo.stsci.edu/coron_spec}}.
 
Coronagraph models can be downloaded directly with \texttt{yippy}'s
\texttt{datasets} module which hosts a catalog of publicly available YIPs that
will be updated as more become available.

No new data were generated or analyzed in support of this research.

\section*{Conflict of Interest (COI) statement}

The authors declare no conflict of interest.


\bibliographystyle{rasti}
\bibliography{biblio} 

@ARTICLE{Robinson2016,
       author = {{Robinson}, Tyler D. and {Stapelfeldt}, Karl R. and {Marley}, Mark S.},
        title = "{Characterizing Rocky and Gaseous Exoplanets with 2 m Class Space-based Coronagraphs}",
      journal = {\pasp},
         year = 2016,
        month = feb,
       volume = {128},
       number = {960},
        pages = {025003},
          doi = {10.1088/1538-3873/128/960/025003},
archivePrefix = {arXiv},
       eprint = {1507.00777},
 primaryClass = {astro-ph.EP},
       adsurl = {https://ui.adsabs.harvard.edu/abs/2016PASP..128b5003R}
}

@article{Lustig-Yaeger2019, doi = {10.21105/joss.01387}, url = {https://doi.org/10.21105/joss.01387}, year = {2019}, publisher = {The Open Journal}, volume = {4}, number = {40}, pages = {1387}, author = {Lustig-Yaeger, Jacob and Robinson, Tyler D. and Arney, Giada}, title = {``coronagraph``: Telescope Noise Modeling for Exoplanets in Python}, journal = {Journal of Open Source Software} }

@article{astropy2013,
        Adsurl = {http://adsabs.harvard.edu/abs/2013A%26A...558A..33A},
        Archiveprefix = {arXiv},
        Author = {{Astropy Collaboration} and {Robitaille}, T.~P. and {Tollerud}, E.~J. and {Greenfield}, P. and {Droettboom}, M. and {Bray}, E. and {Aldcroft}, T. and {Davis}, M. and {Ginsburg}, A. and {Price-Whelan}, A.~M. and {Kerzendorf}, W.~E. and {Conley}, A. and {Crighton}, N. and {Barbary}, K. and {Muna}, D. and {Ferguson}, H. and {Grollier}, F. and {Parikh}, M.~M. and {Nair}, P.~H. and {Unther}, H.~M. and {Deil}, C. and {Woillez}, J. and {Conseil}, S. and {Kramer}, R. and {Turner}, J.~E.~H. and {Singer}, L. and {Fox}, R. and {Weaver}, B.~A. and {Zabalza}, V. and {Edwards}, Z.~I. and {Azalee Bostroem}, K. and {Burke}, D.~J. and {Casey}, A.~R. and {Crawford}, S.~M. and {Dencheva}, N. and {Ely}, J. and {Jenness}, T. and {Labrie}, K. and {Lim}, P.~L. and {Pierfederici}, F. and {Pontzen}, A. and {Ptak}, A. and {Refsdal}, B. and {Servillat}, M. and {Streicher}, O.},
        Doi = {10.1051/0004-6361/201322068},
        Eid = {A33},
        Eprint = {1307.6212},
        Journal = {\aap},
        Month = oct,
        Pages = {A33},
        Primaryclass = {astro-ph.IM},
        Title = {{Astropy: A community Python package for astronomy}},
        Volume = 558,
        Year = 2013}

@ARTICLE{astropy2018,
               author = {{Astropy Collaboration} and {Price-Whelan}, A.~M. and
                 {Sip{\H{o}}cz}, B.~M. and {G{\"u}nther}, H.~M. and {Lim}, P.~L. and
                 {Crawford}, S.~M. and {Conseil}, S. and {Shupe}, D.~L. and
                 {Craig}, M.~W. and {Dencheva}, N. and {Ginsburg}, A. and {Vand
                erPlas}, J.~T. and {Bradley}, L.~D. and {P{\'e}rez-Su{\'a}rez}, D. and
                 {de Val-Borro}, M. and {Aldcroft}, T.~L. and {Cruz}, K.~L. and
                 {Robitaille}, T.~P. and {Tollerud}, E.~J. and {Ardelean}, C. and
                 {Babej}, T. and {Bach}, Y.~P. and {Bachetti}, M. and {Bakanov}, A.~V. and
                 {Bamford}, S.~P. and {Barentsen}, G. and {Barmby}, P. and
                 {Baumbach}, A. and {Berry}, K.~L. and {Biscani}, F. and {Boquien}, M. and
                 {Bostroem}, K.~A. and {Bouma}, L.~G. and {Brammer}, G.~B. and
                 {Bray}, E.~M. and {Breytenbach}, H. and {Buddelmeijer}, H. and
                 {Burke}, D.~J. and {Calderone}, G. and {Cano Rodr{\'\i}guez}, J.~L. and
                 {Cara}, M. and {Cardoso}, J.~V.~M. and {Cheedella}, S. and {Copin}, Y. and
                 {Corrales}, L. and {Crichton}, D. and {D'Avella}, D. and {Deil}, C. and
                 {Depagne}, {\'E}. and {Dietrich}, J.~P. and {Donath}, A. and
                 {Droettboom}, M. and {Earl}, N. and {Erben}, T. and {Fabbro}, S. and
                 {Ferreira}, L.~A. and {Finethy}, T. and {Fox}, R.~T. and
                 {Garrison}, L.~H. and {Gibbons}, S.~L.~J. and {Goldstein}, D.~A. and
                 {Gommers}, R. and {Greco}, J.~P. and {Greenfield}, P. and
                 {Groener}, A.~M. and {Grollier}, F. and {Hagen}, A. and {Hirst}, P. and
                 {Homeier}, D. and {Horton}, A.~J. and {Hosseinzadeh}, G. and {Hu}, L. and
                 {Hunkeler}, J.~S. and {Ivezi{\'c}}, {\v{Z}}. and {Jain}, A. and
                 {Jenness}, T. and {Kanarek}, G. and {Kendrew}, S. and {Kern}, N.~S. and
                 {Kerzendorf}, W.~E. and {Khvalko}, A. and {King}, J. and {Kirkby}, D. and
                 {Kulkarni}, A.~M. and {Kumar}, A. and {Lee}, A. and {Lenz}, D. and
                 {Littlefair}, S.~P. and {Ma}, Z. and {Macleod}, D.~M. and
                 {Mastropietro}, M. and {McCully}, C. and {Montagnac}, S. and
                 {Morris}, B.~M. and {Mueller}, M. and {Mumford}, S.~J. and {Muna}, D. and
                 {Murphy}, N.~A. and {Nelson}, S. and {Nguyen}, G.~H. and
                 {Ninan}, J.~P. and {N{\"o}the}, M. and {Ogaz}, S. and {Oh}, S. and
                 {Parejko}, J.~K. and {Parley}, N. and {Pascual}, S. and {Patil}, R. and
                 {Patil}, A.~A. and {Plunkett}, A.~L. and {Prochaska}, J.~X. and
                 {Rastogi}, T. and {Reddy Janga}, V. and {Sabater}, J. and
                 {Sakurikar}, P. and {Seifert}, M. and {Sherbert}, L.~E. and
                 {Sherwood-Taylor}, H. and {Shih}, A.~Y. and {Sick}, J. and
                 {Silbiger}, M.~T. and {Singanamalla}, S. and {Singer}, L.~P. and
                 {Sladen}, P.~H. and {Sooley}, K.~A. and {Sornarajah}, S. and
                 {Streicher}, O. and {Teuben}, P. and {Thomas}, S.~W. and
                 {Tremblay}, G.~R. and {Turner}, J.~E.~H. and {Terr{\'o}n}, V. and
                 {van Kerkwijk}, M.~H. and {de la Vega}, A. and {Watkins}, L.~L. and
                 {Weaver}, B.~A. and {Whitmore}, J.~B. and {Woillez}, J. and
                 {Zabalza}, V. and {Astropy Contributors}},
                title = "{The Astropy Project: Building an Open-science Project and Status of the v2.0 Core Package}",
              journal = {\aj},
                 year = 2018,
                month = sep,
               volume = {156},
               number = {3},
                  eid = {123},
                pages = {123},
                  doi = {10.3847/1538-3881/aabc4f},
        archivePrefix = {arXiv},
               eprint = {1801.02634},
         primaryClass = {astro-ph.IM},
               adsurl = {https://ui.adsabs.harvard.edu/abs/2018AJ....156..123A}
        }

@ARTICLE{astropy2022,
               author = {{Astropy Collaboration} and {Price-Whelan}, Adrian M. and {Lim}, Pey Lian and {Earl}, Nicholas and {Starkman}, Nathaniel and {Bradley}, Larry and {Shupe}, David L. and {Patil}, Aarya A. and {Corrales}, Lia and {Brasseur}, C.~E. and {N{"o}the}, Maximilian and {Donath}, Axel and {Tollerud}, Erik and {Morris}, Brett M. and {Ginsburg}, Adam and {Vaher}, Eero and {Weaver}, Benjamin A. and {Tocknell}, James and {Jamieson}, William and {van Kerkwijk}, Marten H. and {Robitaille}, Thomas P. and {Merry}, Bruce and {Bachetti}, Matteo and {G{"u}nther}, H. Moritz and {Aldcroft}, Thomas L. and {Alvarado-Montes}, Jaime A. and {Archibald}, Anne M. and {B{'o}di}, Attila and {Bapat}, Shreyas and {Barentsen}, Geert and {Baz{'a}n}, Juanjo and {Biswas}, Manish and {Boquien}, M{'e}d{'e}ric and {Burke}, D.~J. and {Cara}, Daria and {Cara}, Mihai and {Conroy}, Kyle E. and {Conseil}, Simon and {Craig}, Matthew W. and {Cross}, Robert M. and {Cruz}, Kelle L. and {D'Eugenio}, Francesco and {Dencheva}, Nadia and {Devillepoix}, Hadrien A.~R. and {Dietrich}, J{"o}rg P. and {Eigenbrot}, Arthur Davis and {Erben}, Thomas and {Ferreira}, Leonardo and {Foreman-Mackey}, Daniel and {Fox}, Ryan and {Freij}, Nabil and {Garg}, Suyog and {Geda}, Robel and {Glattly}, Lauren and {Gondhalekar}, Yash and {Gordon}, Karl D. and {Grant}, David and {Greenfield}, Perry and {Groener}, Austen M. and {Guest}, Steve and {Gurovich}, Sebastian and {Handberg}, Rasmus and {Hart}, Akeem and {Hatfield-Dodds}, Zac and {Homeier}, Derek and {Hosseinzadeh}, Griffin and {Jenness}, Tim and {Jones}, Craig K. and {Joseph}, Prajwel and {Kalmbach}, J. Bryce and {Karamehmetoglu}, Emir and {Ka{l}uszy{'n}ski}, Miko{l}aj and {Kelley}, Michael S.~P. and {Kern}, Nicholas and {Kerzendorf}, Wolfgang E. and {Koch}, Eric W. and {Kulumani}, Shankar and {Lee}, Antony and {Ly}, Chun and {Ma}, Zhiyuan and {MacBride}, Conor and {Maljaars}, Jakob M. and {Muna}, Demitri and {Murphy}, N.~A. and {Norman}, Henrik and {O'Steen}, Richard and {Oman}, Kyle A. and {Pacifici}, Camilla and {Pascual}, Sergio and {Pascual-Granado}, J. and {Patil}, Rohit R. and {Perren}, Gabriel I. and {Pickering}, Timothy E. and {Rastogi}, Tanuj and {Roulston}, Benjamin R. and {Ryan}, Daniel F. and {Rykoff}, Eli S. and {Sabater}, Jose and {Sakurikar}, Parikshit and {Salgado}, Jes{'u}s and {Sanghi}, Aniket and {Saunders}, Nicholas and {Savchenko}, Volodymyr and {Schwardt}, Ludwig and {Seifert-Eckert}, Michael and {Shih}, Albert Y. and {Jain}, Anany Shrey and {Shukla}, Gyanendra and {Sick}, Jonathan and {Simpson}, Chris and {Singanamalla}, Sudheesh and {Singer}, Leo P. and {Singhal}, Jaladh and {Sinha}, Manodeep and {Sip{H{o}}cz}, Brigitta M. and {Spitler}, Lee R. and {Stansby}, David and {Streicher}, Ole and {{\v{S}}umak}, Jani and {Swinbank}, John D. and {Taranu}, Dan S. and {Tewary}, Nikita and {Tremblay}, Grant R. and {Val-Borro}, Miguel de and {Van Kooten}, Samuel J. and {Vasovi{'c}}, Zlatan and {Verma}, Shresth and {de Miranda Cardoso}, Jos{'e} Vin{'i}cius and {Williams}, Peter K.~G. and {Wilson}, Tom J. and {Winkel}, Benjamin and {Wood-Vasey}, W.~M. and {Xue}, Rui and {Yoachim}, Peter and {Zhang}, Chen and {Zonca}, Andrea and {Astropy Project Contributors}},
                title = "{The Astropy Project: Sustaining and Growing a Community-oriented Open-source Project and the Latest Major Release (v5.0) of the Core Package}",
              journal = {\apj},
                 year = 2022,
                month = aug,
               volume = {935},
               number = {2},
                  eid = {167},
                pages = {167},
                  doi = {10.3847/1538-4357/ac7c74},
        archivePrefix = {arXiv},
               eprint = {2206.14220},
         primaryClass = {astro-ph.IM},
               adsurl = {https://ui.adsabs.harvard.edu/abs/2022ApJ...935..167A}
        }

@ARTICLE{2025arXiv250218556S,
       author = {{Stark}, Christopher C. and {Steiger}, Sarah and {Tokadjian}, Armen and {Savransky}, Dmitry and {Belikov}, Rus and {Chen}, Pin and {Krist}, John and {Macintosh}, Bruce and {Morgan}, Rhonda and {Pueyo}, Laurent and {Sirbu}, Dan and {Stapelfeldt}, Karl},
        title = "{Cross-Model Validation of Coronagraphic Exposure Time Calculators for the Habitable Worlds Observatory: A Report from the Exoplanet Science Yield sub-Working Group}",
      journal = {arXiv e-prints},
         year = 2025,
        month = feb,
          eid = {arXiv:2502.18556},
        pages = {arXiv:2502.18556},
          doi = {10.48550/arXiv.2502.18556},
archivePrefix = {arXiv},
       eprint = {2502.18556},
 primaryClass = {astro-ph.IM},
       adsurl = {https://ui.adsabs.harvard.edu/abs/2025arXiv250218556S}
}

@article{latouf_determining_2026,
  title = {Determining the {{Detectability}} of {{H2O}} with {{Photometric Observations Using Bayesian Analysis}} for {{Remote Biosignature Identification}} on Exo-{{Earths}} ({{BARBIE}})},
  author = {Latouf, Natasha and Stark, Chris and Mandell, Avi M. and Kofman, Vincent},
  year = 2026,
  month = jan,
  journal = {The Astronomical Journal},
  volume = {171},
  number = {2},
  pages = {74},
  publisher = {The American Astronomical Society},
  issn = {1538-3881},
  doi = {10.3847/1538-3881/ae29b1},
  urldate = {2026-04-13},
  langid = {english}
}

@article{spohn_yippy_2026,
    author = {Spohn, Corey and Stark, Christopher C. and Redmond, Susan and Juanola-Parramon, Roser},
    title = {{yippy}: Standardized Yield Input Package Processing and Fourier-Domain {PSF} Interpolation for {HWO} Coronagraphs},
    journal = {RAS Techniques and Instruments},
    year = {2026},
    note = {in prep},
}

@techreport{starkStandardizedCoronagraph,
  title = {Standardized {{Coronagraph Parameters}} for {{Input}} into {{Yield Calculations}}},
  author = {Stark, Christopher and Krist, John},
  year = 2019,
  month = feb,
  langid = {english},
  institution = {NASA}
}

@inproceedings{belikov_coronagraph_2024,
  title = {Coronagraph Design Survey for Future Exoplanet Direct Imaging Space Missions},
  booktitle = {Space {{Telescopes}} and {{Instrumentation}} 2024: {{Optical}}, {{Infrared}}, and {{Millimeter Wave}}},
  author = {Belikov, Ruslan and Stark, Christopher and Siegler, Nick and Por, Emiel and Mennesson, Bertrand and Redmond, Susan and Chen, Pin and Fogarty, Kevin and Guyon, Olivier and {Juanola-Parramon}, Roser and Kasdin, Jeremy and Krist, John and Mawet, Dimitri and Morgan, Rhonda and Mejia Prada, Camilo and Pueyo, Laurent and Ruane, Garreth and Sirbu, Dan and Stapelfeldt, Karl and Trauger, John and Zimmerman, Neil and Alagao, Mary Angelie M. and Carlotti, Alex and Chafi, Jamal and Doleman, David and {Gersh-Range}, Jessica and Konig, Lorenzo and Leboulleux, Lucille and Moody, Dwight and Riggs, A. J. and Serabyn, Eugene and Snik, Frans and Wallace, Kent},
  year = 2024,
  month = aug,
  volume = {13092},
  pages = {1309266},
  doi = {10.1117/12.3020614},
  urldate = {2025-09-17}
}

@book{NAP26141,
    title = {Pathways to Discovery in Astronomy and Astrophysics for the 2020s},
	author = {{National Academies of Sciences, Engineering, and Medicine}},
	doi = {10.17226/26141},
	publisher = {The National Academies Press},
	year = 2021,
	}

@inproceedings{Feinberg2024,
author = {Lee Feinberg and John Ziemer and Megan Ansdell and Julie Crooke and Courtney Dressing and Bertrand Mennesson and John O'Meara and Joshua Pepper and Aki Roberge},
title = {{The Habitable Worlds Observatory engineering view: status, plans, and opportunities}},
volume = {13092},
booktitle = {Space Telescopes and Instrumentation 2024: Optical, Infrared, and Millimeter Wave},
editor = {Laura E. Coyle and Shuji Matsuura and Marshall D. Perrin},
organization = {International Society for Optics and Photonics},
publisher = {SPIE},
pages = {130921N},
year = {2024},
doi = {10.1117/12.3018328},
URL = {https://doi.org/10.1117/12.3018328}
}

@ARTICLE{Feinberg2026,
       author = {{Feinberg}, Lee D. and {Sitarski}, Breann N. and {McElwain}, Michael W. and {Arney}, Giada and {Baker}, Caleb and {Bolcar}, Matthew D. and {Levine}, Marie and {Liu}, Alice and {Mennesson}, Bertrand and {Roberge}, Aki and {Smith}, J. Scott and {Zhao}, Feng and {Ziemer}, John},
        title = "{Habitable Worlds Observatory's Concept and Technology Maturation: Initial Feasibility and Trade Space Exploration}",
      journal = {arXiv e-prints},
         year = 2026,
        month = jan,
          eid = {arXiv:2601.11803},
        pages = {arXiv:2601.11803},
archivePrefix = {arXiv},
       eprint = {2601.11803},
 primaryClass = {astro-ph.IM},
       adsurl = {https://ui.adsabs.harvard.edu/abs/2026arXiv260111803F}
}

@ARTICLE{2014ApJ...795..122S,
       author = {{Stark}, Christopher C. and {Roberge}, Aki and {Mandell}, Avi and {Robinson}, Tyler D.},
        title = "{Maximizing the ExoEarth Candidate Yield from a Future Direct Imaging Mission}",
      journal = {The Astrophysical Journal},
         year = 2014,
        month = nov,
       volume = {795},
       number = {2},
          eid = {122},
        pages = {122},
          doi = {10.1088/0004-637X/795/2/122},
archivePrefix = {arXiv},
       eprint = {1409.5128},
 primaryClass = {astro-ph.SR},
       adsurl = {https://ui.adsabs.harvard.edu/abs/2014ApJ...795..122S}
}

@ARTICLE{2019JATIS...5b4009S,
       author = {{Stark}, Christopher C. and {Belikov}, Rus and {Bolcar}, Matthew R. and {Cady}, Eric and {Crill}, Brendan P. and {Ertel}, Steve and {Groff}, Tyler and {Hildebrandt}, Sergi and {Krist}, John and {Lisman}, P. Douglas and {Mazoyer}, Johan and {Mennesson}, Bertrand and {Nemati}, Bijan and {Pueyo}, Laurent and {Rauscher}, Bernard J. and {Riggs}, A.~J. and {Ruane}, Garreth and {Shaklan}, Stuart B. and {Sirbu}, Dan and {Soummer}, Remi and {Laurent}, Kathryn St. and {Zimmerman}, Neil},
        title = "{ExoEarth yield landscape for future direct imaging space telescopes}",
      journal = {Journal of Astronomical Telescopes, Instruments, and Systems},
         year = 2019,
        month = apr,
       volume = {5},
          eid = {024009},
        pages = {024009},
          doi = {10.1117/1.JATIS.5.2.024009},
archivePrefix = {arXiv},
       eprint = {1904.11988},
 primaryClass = {astro-ph.EP},
       adsurl = {https://ui.adsabs.harvard.edu/abs/2019JATIS...5b4009S}
}

@ARTICLE{2026arXiv260106342S,
       author = {{Steiger}, Sarah and {Chen}, Pin and {Pueyo}, Laurent},
        title = "{Incorporating Wavefront Error, Wavefront Sensing and Control, and Sensitivities into Exposure Time Calculations for Future Space Missions with the Error Budget Software (EBS)}",
      journal = {arXiv e-prints},
         year = 2026,
        month = jan,
          eid = {arXiv:2601.06342},
        pages = {arXiv:2601.06342},
          doi = {10.48550/arXiv.2601.06342},
archivePrefix = {arXiv},
       eprint = {2601.06342},
 primaryClass = {astro-ph.IM},
       adsurl = {https://ui.adsabs.harvard.edu/abs/2026arXiv260106342S}
}

@ARTICLE{2026AJ....171..228A,
       author = {{Alei}, Eleonora and {Mandell}, Avi M. and {Currie}, Miles H. and {Roberge}, Aki and {Stark}, Christopher C. and {Payne}, Allison and {Kofman}, Vincent and {Villanueva}, Geronimo L. and {Hu}, Renyu and {Young}, Amber V.},
        title = "{Multibandpass Photometry for Exoplanet Atmosphere Reconnaissance (MPEAR) with the Habitable Worlds Observatory (HWO). I. Differentiating Earth from Neptunes during Discovery}",
      journal = {\aj},
         year = 2026,
        month = apr,
       volume = {171},
       number = {4},
          eid = {228},
        pages = {228},
          doi = {10.3847/1538-3881/ae4597},
archivePrefix = {arXiv},
       eprint = {2512.05279},
 primaryClass = {astro-ph.EP},
       adsurl = {https://ui.adsabs.harvard.edu/abs/2026AJ....171..228A}
}

@ARTICLE{1998A&AS..127....1L,
       author = {{Leinert}, Ch. and {Bowyer}, S. and {Haikala}, L.~K. and {Hanner}, M.~S. and {Hauser}, M.~G. and {Levasseur-Regourd}, A.-Ch. and {Mann}, I. and {Mattila}, K. and {Reach}, W.~T. and {Schlosser}, W. and {Staude}, H.~J. and {Toller}, G.~N. and {Weiland}, J.~L. and {Weinberg}, J.~L. and {Witt}, A.~N.},
        title = "{The 1997 reference of diffuse night sky brightness}",
      journal = {\aaps},
         year = 1998,
        month = jan,
       volume = {127},
        pages = {1-99},
          doi = {10.1051/aas:1998105},
       adsurl = {https://ui.adsabs.harvard.edu/abs/1998A&AS..127....1L}
}

@article{nemati2023analytical,
  title={Analytical performance model and error budget for the Roman coronagraph instrument},
  author={Nemati, Bijan and Krist, John and Poberezhskiy, Ilya and Kern, Brian},
  journal={Journal of Astronomical Telescopes, Instruments, and Systems},
  volume={9},
  number={3},
  pages={034007--034007},
  year={2023},
  publisher={Society of Photo-Optical Instrumentation Engineers}
}

@ARTICLE{Biancalani2025,
       author = {{Biancalani}, Enrico and {Balaban}, Edward and {Belikov}, Ruslan and {Bendek}, Eduardo and {Frumkin}, Valeri and {Gabay}, Israel and {Gao}, Guangjun and {Gong}, Qian and {Gregg}, Christine and {Groff}, Tyler and {Howard}, Joseph and {Luria}, Omer and {McElwain}, Michael and {Mundy}, Lee and {Ticknor}, Rachel and {Veilleux}, Sylvain and {Zimmerman}, Neil},
        title = "{Optical Design Pathways to Fluidic Space-Assembled Reflectors \& Dual-Configuration Spectrographs for Characterizing Exo-Earths}",
      journal = {arXiv e-prints},
         year = 2025,
        month = oct,
          eid = {arXiv:2510.02479},
        pages = {arXiv:2510.02479},
          doi = {10.48550/arXiv.2510.02479},
archivePrefix = {arXiv},
       eprint = {2510.02479},
 primaryClass = {astro-ph.IM},
       adsurl = {https://ui.adsabs.harvard.edu/abs/2025arXiv251002479B}
}

@ARTICLE{2026arXiv260426925G,
       author = {{Gilbert-Janizek}, Samantha and {Lustig-Yaeger}, Jacob and {Krissansen-Totton}, Joshua},
        title = "{The effect of spectral resolution on biosignature detection via reflected light observations of the Earth through time}",
      journal = {arXiv e-prints},
         year = 2026,
        month = apr,
          eid = {arXiv:2604.26925},
        pages = {arXiv:2604.26925},
          doi = {10.48550/arXiv.2604.26925},
archivePrefix = {arXiv},
       eprint = {2604.26925},
 primaryClass = {astro-ph.EP},
       adsurl = {https://ui.adsabs.harvard.edu/abs/2026arXiv260426925G}
}

@ARTICLE{2026arXiv260714329C,
       author = {{Currie}, Miles H. and {Stark}, Christopher C. and {Alei}, Eleonora and {Roberge}, Aki},
        title = "{The exozodi spectral effect: Residual habitable zone dust may bias exoEarth characterization}",
      journal = {arXiv e-prints},
         year = 2026,
        month = jul,
          eid = {arXiv:2607.14329},
        pages = {arXiv:2607.14329},
          doi = {10.48550/arXiv.2607.14329},
archivePrefix = {arXiv},
       eprint = {2607.14329},
 primaryClass = {astro-ph.EP},
       adsurl = {https://ui.adsabs.harvard.edu/abs/2026arXiv260714329C}
}

@software{2017ascl.soft06010S,
       author = {{Savransky}, Dmitry and {Delacroix}, Christian and {Garrett}, Daniel},
        title = "{EXOSIMS: Exoplanet Open-Source Imaging Mission Simulator}",
 howpublished = {Astrophysics Source Code Library, record ascl:1706.010},
         year = 2017,
        month = jun,
          eid = {ascl:1706.010},
archivePrefix = {ascl},
       eprint = {1706.010},
       adsurl = {https://ui.adsabs.harvard.edu/abs/2017ascl.soft06010S}
}

@software{pyEDITHzenodo,
  author       = {Eleonora Alei and
                  Miles Currie and
                  Corey Spohn},
  title        = {HabitableWorldsObservatory/pyEDITH},
  month        = aug,
  year         = 2026,
  publisher    = {Zenodo},
  doi          = {10.5281/zenodo.17917471},
  url          = {https://doi.org/10.5281/zenodo.17917471},
  howpublished = {Software, Zenodo, doi:10.5281/zenodo.17917471},
}

@article{Nemati2014,
title = {Detector Selection for the {{WFIRST-AFTA}} Coronagraph Integral Field Spectrograph},
author = {Nemati, Bijan},
year = 2014,
month = aug,
journal = {SPIE},
volume = {9143},
pages = {196--201},
publisher = {SPIE},
doi = {10.1117/12.2060321},
urldate = {2021-10-20},
}

@article{Garrett2017b,
title = {A {{Simple Depth}} of {{Search Metric}} for {{Exoplanet Imaging Surveys}}},
author = {Garrett, Daniel and Savransky, Dmitry and Macintosh, Bruce},
year = 2017,
journal = {The Astronomical Journal},
volume = {154},
number = {2},
eprint = {1706.06132},
pages = {47},
publisher = {IOP Publishing},
issn = {1538-3881},
doi = {10.3847/1538-3881/aa78f6},
}

@article{10.1117/1.JATIS.12.4.041011,
  title = {Apodized Vortex Coronagraph for the Habitable Worlds Observatory},
  author = {Redmond, Susan F. and Mawet, Dimitri and {Bertrou-Cantou}, Arielle and Ruane, Garreth and {Llop-Sayson}, Jorge},
  year = 2026,
  journal = {Journal of Astronomical Telescopes, Instruments, and Systems},
  volume = {12},
  number = {4},
  pages = {041011},
  publisher = {SPIE},
  doi = {10.1117/1.JATIS.12.4.041011}
}




\appendix

\section{Example input file}

In \autoref{fig:inputfile} we report an example of a template input file that \texttt{pyEDITH} can read from command line.
\begin{figure*}
\begin{lstlisting}
;This is an input file for pyEDITH
;Comments are preceded with a # or a ;
;In the comments below, ( ) indicates the units and { } indicates the data format

;Variable names cannot be changed
;Arrays are encased by [ ] and entries are separated by ,
;Scalars cannot be encased in [ ]--scalars are not the same as one-element arrays
;Spaces do not matter, empty lines do not matter

;--------------------------------
;------- REFERENCE LAMBDA -------
;--------------------------------

;--- OBSERVATIONAL PARAMETERS ---
wavelength = 0.5                    ; wavelength (micron)
snr = 7.0                           ; signal-to-noise ratio at desired wavelength {same length as wavelength} 
CRb_multiplier  = 2.                ; factor to multiply to remove background
psf_trunc_ratio = 0.35              ; truncation ratio of the PSF 

;--- STAR PARAMETERS ---
distance = 14.8                     ; distance to star (pc)
FstarV_10pc = 1.244e+02             ; flux in the V band (ph/s/cm^2/nm)
Fstar_10pc = 1.128e+02              ; flux at desired wavelength {same length as wavelength} (ph/s/cm^2/nm)
stellar_radius = 0.95               ; stellar radius (R_sun)
nzodis = 3.0                        ; amount of exozodi around target star ("zodis")
ra = 236.0075773682300              ; right ascension of target star used to estimate zodi (deg) 
dec = 02.5151668316500              ; declination of target star used to estimate zodi (deg)

;--- PLANET PARAMETERS ---
separation = 0.062                  ; separation of planet (arcsec)
Fp/Fs =  6.3e-8                     ; planet contrast at desired wavelength {same length as wavelength} 

;--- OBSERVATORY PARAMETERS ----
observatory_preset = EAC1       
observing_mode =  IMAGER

;--------------------------------
;------- SECONDARY LAMBDA -------
;--------------------------------

;--- OBSERVATIONAL PARAMETERS ---
secondary_wavelength = 1.0          ; wavelength (micron)

;--- STAR PARAMETERS ---
secondary_Fstar_10pc = 1.102e+02    ; flux at secondary wavelength {same length as wavelength} (ph/s/cm^2/nm)

;--- PLANET PARAMETERS ---
secondary_Fp/Fs = 6.4e-8            ; planet contrast at secondary wavelength {same length as wavelength} 

\end{lstlisting}
\caption{Example input file for \texttt{pyEDITH}{, for the calculation of the exposure time necessary to achieve S/N=7 at 500 nm for a bright planet at 0.062 arcsec ($\approx4.78\ \lambda/D$) separation and 14.8 pc distance around a fiducial star similar to HIP 77052, one of the fiducial stars used in the ETC comparison performed by \citet{2025arXiv250218556S}. We also show the optional ``secondary wavelength'' section, which instructs \texttt{pyEDITH} to first calculate the exposure time at the reference wavelength and then use that time to evaluate the S/N at the secondary wavelength (1 $\mu$m in this case). This approach has been used in \citet{2026AJ....171..228A}. }}\label{fig:inputfile}
\end{figure*}


\bsp	
\label{lastpage}
\end{document}